\documentclass[fleqn,usenatbib]{mnras}

\usepackage{newtxtext,newtxmath}
\usepackage[T1]{fontenc}
\usepackage{soul}

\usepackage{hyperref}

\DeclareRobustCommand{\VAN}[3]{#2}
\let\VANthebibliography\thebibliography
\def\thebibliography{\DeclareRobustCommand{\VAN}[3]{##3}\VANthebibliography}

\usepackage{graphicx}	% Including figure files
\usepackage{amsmath}	% Advanced maths commands
\usepackage[dvipsnames]{xcolor}

\newcommand{\Mpc}{$h^{-1}$Mpc}
\newcommand{\cov}{{\rm cov}}
\newcommand{\diag}{{\rm diag}}
\title[The 2-point correlation function covariance with fewer mocks]{The 2-point correlation function covariance with fewer mocks}

\author[S. Trusov et al. ]{Svyatoslav Trusov$^{1}$\thanks{E-mail: strusov@lpnhe.in2p3.fr},
Pauline Zarrouk$^{1}$,
Shaun Cole$^{2}$,
Peder Norberg$^{2,3}$, 
Cheng Zhao$^{4}$, 
\newauthor
Jessica Nicole Aguilar$^{5}$,
Steven Ahlen$^{6}$, 
David Brooks$^{7}$, 
Axel de la Macorra$^{8}$, 
Peter Doel$^{7}$, 
Andreu Font-Ribera$^{9}$, 
\newauthor
Klaus Honscheid$^{10,}$$^{11,}$$^{12}$, 
Theodore Kisner$^{5}$, 
Martin Landriau$^{5}$, 
Christophe Magneville$^{13}$, 
Ramon Miquel$^{9,}$$^{14}$, 
\newauthor
Jundan Nie$^{15}$, 
Claire Poppett$^{5,}$$^{16,}$$^{17}$, 
Michael Schubnell$^{18}$, 
Gregory Tarlé$^{18}$, 
Zhimin Zhou$^{15}$ 
\\
$^{1}$ Sorbonne Universit\'e, Universit\'e Paris Diderot, Sorbonne Paris Cit\'e, CNRS,
Laboratoire de Physique Nucléaire et de Hautes Energies (LPNHE).\\
4 place Jussieu, F-75252, Paris
Cedex 5, France\\
$^{2}$ Institute for Computational Cosmology, Department of Physics, Durham University, South Road, Durham DH1 3LE, UK.\\
$^{3}$ Centre for Extragalactic Astronomy, Department of Physics, Durham University, South Road, Durham DH1 3LE, UK.\\
$^{4}$  Institute of Physics, Laboratory of Astrophysics, École Polytechnique Fédérale de Lausanne (EPFL), Observatoire de Sauverny, CH-1290 Versoix, Switzerland. \\
$^{5}$ Lawrence Berkeley National Laboratory, 1 Cyclotron Road, Berkeley, CA 94720, USA\\
$^{6}$ Physics Dept., Boston University, 590 Commonwealth Avenue, Boston, MA 02215, USA \\
$^{7}$ Department of Physics \& Astronomy, University College London, Gower Street, London, WC1E 6BT, UK  \\
$^{8}$ Instituto de F\'{\i}sica, Universidad Nacional Aut\'{o}noma de M\'{e}xico,  Cd. de M\'{e}xico  C.P. 04510,  M\'{e}xico \\
$^{9}$ Institut de F\'{i}sica d’Altes Energies (IFAE), The Barcelona Institute of Science and Technology, Campus UAB, 08193 Bellaterra Barcelona, Spain\\
$^{10}$ Center for Cosmology and AstroParticle Physics, The Ohio State University, 191 West Woodruff Avenue, Columbus, OH 43210, USA \\
$^{11}$ Department of Physics, The Ohio State University, 191 West Woodruff Avenue, Columbus, OH 43210, USA \\
$^{12}$ The Ohio State University, Columbus, 43210 OH, USA \\
$^{13}$ IRFU, CEA, Universit\'{e} Paris-Saclay, F-91191 Gif-sur-Yvette, France \\
$^{14}$ Instituci\'{o} Catalana de Recerca i Estudis Avan\c{c}ats, Passeig de Llu\'{\i}s Companys, 23, 08010 Barcelona, Spain \\
$^{15}$ National Astronomical Observatories, Chinese Academy of Sciences, A20 Datun Rd., Chaoyang District, Beijing, 100012, P.R. China \\
$^{16}$ Space Sciences Laboratory, University of California, Berkeley, 7 Gauss Way, Berkeley, CA  94720, USA \\
$^{17}$ University of California, Berkeley, 110 Sproul Hall \#5800 Berkeley, CA 94720, USA\\
$^{18}$ Department of Physics, University of Michigan, Ann Arbor, MI 48109, USA \\
}

\date{Accepted XXX. Received YYY; in original form ZZZ}

\pubyear{2023}

\begin{document}
\label{firstpage}
\pagerange{\pageref{firstpage}--\pageref{lastpage}}
\maketitle

\begin{abstract}
We present \texttt{FitCov} an approach for accurate estimation of the covariance of 2-point correlation functions that requires fewer mocks than the standard mock-based covariance. This can be achieved by dividing a set of mocks into jackknife regions and fitting the correction term first introduced in Mohammad \& Percival (2022), such that the mean of the jackknife covariances corresponds to the one from the mocks. This extends the model beyond the shot-noise limited regime, allowing it to be used for denser samples of galaxies. We test the performance of our fitted jackknife approach, both in terms of accuracy and precision, using lognormal mocks with varying densities and approximate EZmocks mimicking the DESI LRG and ELG samples in the redshift range of $z = [0.8,1.1]$. We find that the Mohammad-Percival correction produces a bias in the 2-point correlation function covariance matrix that grows with number density and that our fitted jackknife approach does not. We also study the effect of the covariance on the uncertainty of cosmological parameters by performing a full-shape analysis. We demonstrate that our fitted jackknife approach based on 25 mocks can recover unbiased and as precise cosmological parameters as the ones obtained from a covariance matrix based on 1000 or 1500 mocks, while the Mohammad-Percival correction produces uncertainties that are twice as large. The number of mocks required to obtain an accurate estimation of the covariance for the 2-point correlation function is therefore reduced by a factor of 40-60. The \texttt{FitCov} code that accompanies this paper is available at \href{https://github.com/theonefromnowhere/FitCov}{this GitHub repository}.
\end{abstract}


%%%%%%%%%%%%%%%%%%%%%%%%%%%%%%%%%%%%%%%%%%%%%%%%%%

%%%%%%%%%%%%%%%%% BODY OF PAPER %%%%%%%%%%%%%%%%%%

\section{Introduction}

A new generation of cosmological surveys
such as Dark Energy Spectroscopic Instrument (\citealt{desi}, \citealt{DESI2022}) 
have started taking data and even more will in the coming years with e.g.\ the 
start of operations of Euclid \citep{Laureijs2011} and the Vera Rubin Observatory \citep{lsst}. Therefore, it is becoming vital to develop methods for deriving covariance matrices in order to estimate the uncertainties on the cosmological parameters of interest. 

Existing methods of evaluating the covariance matrix that quantifies the errors on the galaxy 2-point correlation function of galaxy redshift surveys can be separated into three different categories: mock-based, analytic and internal, each best suited to different scenarios. 
Mock-based covariance matrices are built from a large suite of numerical simulations, ‘mock’ catalogues, that mimic the properties of the cosmological surveys with high fidelity. These mocks need to be i) accurate in the sense that they have to reproduce the two- and higher-point statistics with limited biases and ii) numerous in order to avoid sample variance, which introduces noise in the covariance matrices that could bias the inferred parameter uncertainties (e.g. \citealt{Dawson2013}, \citealt{Percival2014}).

Analytic approaches provide expectation values of the large-scale structure statistics directly and are much less computationally expensive. However, that requires a description of the non-Gaussian terms that enter the four-point correlation function, which is needed to compute the covariance of the two-point correlation function. Accurate modelling of the non-linear gravitational evolution, galaxy bias, redshift-space distortions and shot noise is thus a challenge to compute analytic covariance matrices. The modelling usually relies on Perturbation Theory (PT) which limits the domain of accuracy to the quasi-linear regime when the density perturbations remain small compared to unity. Moreover, one also needs to account for survey geometry and window function effects. Recent progress in this direction has been made to develop codes for the power spectrum (CovaPT, \citealt{Wadekar1}). Additionally, we can mention semi-analytic approaches, which use the data to calibrate themselves, for example, RascalC code (\citealt{RascalC1}, \citealt{RascalC2}).

Finally, data-based or internal methods, such as jackknife and bootstrap, are often used especially when large sets of mocks are not available. They consist in resampling the survey data by slicing the original data into sub-samples and weighting these sub-samples following specific prescriptions. In the standard jackknife approach, for a given jackknife realisation $i$, the sub-samples have unit weight except the sub-sample indexed $i$ that is weighted 0, hence this approach is also called ‘delete-one’ jackknife resampling. Internal resampling methods do not rely on any assumption about the underlying gravity model and are thus less sensitive to unknown physics. However, they can lack precision and suffer from biases, as discussed in \cite{Nordberg}, \cite{Friedrich2016} and \cite{Favelo2020}. One fundamental deficiency of all internal covariance estimators is the large-scale bias coming from a lack of a proper estimation of the super-sample covariance \citep{Lacasa2017}, which is due to the lack of modes larger than the survey size.
Recently, a correction to the standard jackknife resampling method was proposed in \cite{Mohammad_Percival} which consists in introducing a different weighting scheme for the cross-pairs than for the auto-pairs, where the auto-pairs are made up of objects that lie in the same sub-sample and cross-pairs of two objects that reside in two distinct sub-samples. Indeed, the choice of assigning weights to pairs of objects is arbitrary and \cite{Mohammad_Percival} tested different prescriptions. They found that by adjusting the weighting of the pairs that compose the estimates of the two-point correlation function, they were able to provide more accurate estimates of the variance than the standard jackknife. However, it remains an internal estimator, with the associated characteristic fundamental problems such as super-sample covariance.

In this work, we follow a similar methodology but propose to go beyond that work by i) considering some cross-pairs that were neglected in both the standard jackknife and the jackknife method with Mohammad-Percival correction, ii) fitting the appropriate weighting scheme to a mock-based covariance built from a smaller number of mocks than for traditional mock-based approach. The paper's outline is as follows: in Section~\ref{sec:intro-cov} we review the formalism associated with the standard jackknife resampling method and the correction proposed in \cite{Mohammad_Percival}. We introduce there the formalism of our proposed hybrid approach, which performance on mocks is presented in Section~\ref{sec:test-mocks} and compared with the original correction for jackknife and with mock-based method for estimating the covariance matrix. We conclude and discuss further prospects in Section~\ref{sec:conclusions}.

\section{Covariance estimators}
\label{sec:intro-cov}

In the present paper, we work in configuration space. We use the Landy-Szalay estimator with double-counting assumed, \citep{LandySzalay}, which can be written as:
\begin{equation}
    \xi(s,\mu) = \frac{DD(s,\mu) - 2DR(s,\mu) + RR(s,\mu)}{RR(s,\mu)}
    \label{eq:Landy_Szalay}
\end{equation}
where $s$ is the redshift space separation of a pair of galaxies, $\mu$ is the cosine of the angle between the separation vector and the line of sight, $\xi(s,\mu)$ is the 2-point correlation function in redshift space, $DD(s,\mu)$ are the binned auto-pair counts of the data catalogue, $RR(s,\mu)$ are the binned pair counts computed from a matching random catalogue, and $DR(s,\mu)$ are the binned cross-pair counts between the random and the data catalogue. All pair counts are assumed to be suitably normalised in Eq.~\ref{eq:Landy_Szalay}.

The 2-point correlation function can be decomposed into Legendre multipoles defined as:

\begin{equation}
    \xi_{\ell}(s) = (2\ell +1)\int^1_0 \xi(s, \mu) L_{\ell}(\mu)\, d\mu,
\end{equation}
where $\ell$ is the order of the multipole, and $L_{\ell}(\mu)$ are the Legendre polynomials.

\subsection{Covariance from data or data-like mocks}

Cosmological simulations can be divided into two categories: i) precise and expensive computationally N-body simulations, which are known to treat properly non-linear gravitational evolution; ii) less accurate approximate mock methods, such as BAM \citep{bam}, COLA \citep{cola}, EZmock \citep{ez_mocks}, \citep{ez_mocks_eboss}, FastPM \citep{fast_pm}, GLAM \citep{glam}, lognormal, PATCHY \citep{patchy} etc. They can provide a good covariance for scales $>10$~\Mpc, but small-scale clustering is not properly resolved.

Assuming a survey with $N_{\rm m}$ mocks, the covariance matrix of the 2-point correlation function is defined as:

\begin{equation}
    C_{ij}=\frac{1}{N_{\rm m}-1}\sum^{N_{\rm m}}_{k=1} \left[ \xi^{[k]}_i - \langle \xi_i \rangle \right] \left[ \xi^{[k]}_j - \langle \xi_j \rangle \right],
    \label{eqn:cov_eq}
\end{equation}
where $\xi^{[k]}_i$ is the $i^{\rm th}$ bin of correlation function of the $k^{\rm th}$ mock, and $\langle\xi_i\rangle$ is the mean over the $N_{\rm m}$ mocks of the $i^{\rm th}$ bin of the correlation function.

However, for some subsets of modern surveys, like the DESI Bright Galaxy Survey (BGS), the number of galaxies and their number density sometimes becomes so large, that even these approximate methods become expensive computationally, posing a problem.

\subsection{Jackknife covariance}

Jackknife is a data resampling approach that involves creating multiple sub-samples of the same dataset by systematically excluding regions of the data. When applied to the cosmological surveys, the footprint is divided into regions of similar area and it is these that are systematically excluded to make the multiple sub-samples.
%by splitting the survey with straight line cuts in RA and then Dec such that each region contains the same number of points in the random catalogue.

This approach has the advantage of making no assumptions regarding non-linear evolution and non-standard physics,  and at the same time is extremely cheap from the computational perspective, as it does not require expensive production of thousands of mocks. Assuming we have cut our dataset into $N_{\rm jk}$ pieces, the covariance matrix is:

\begin{equation}
    C_{ij}=\frac{N_{\rm jk}-1}{N_{\rm jk}}\sum^{N_{\rm jk}}_{k=1} \left[ \xi^{[k]}_i - \langle \xi_i \rangle \right] \left[ \xi^{[k]}_j - \langle \xi_j \rangle \right],
    \label{eqn:jk_eq}
\end{equation}
where $\xi^{[k]}_i$ is the $i^{\rm th}$  bin of the correlation function of the $k^{\rm th}$  jackknife region, and $\langle\xi_i\rangle$ is its mean over all the $N_{\rm jk}$ jackknife regions.
The coefficient on the right-hand side is larger than the corresponding factor in Eq.~\ref{eqn:cov_eq} as it compensates for the reduction in the covariance due to the overlap between the subsamples.

In practice, we consider the galaxy 2-point correlation function and the $DD$, $DR$ and $RR$ pair counts mentioned in the Landy-Szalay estimator defined in Eq.~\eqref{eq:Landy_Szalay}.

\subsubsection{Standard approach}

We will assume the number of sub-samples is $N_{\rm jk}$ and work in terms of pair counts rather than correlation functions.
 For simplicity, we will denote as $AA_k$ the auto-counts that are contributed by pairs of galaxies that both reside in the $k^{\rm th}$ area of the survey (the areas that are systematically excluded to form the jackknife sub-samples) and $CC_k$ the cross-counts between galaxies in this $k^{\rm th}$ area and those in the jackknife sub-sample that is made by excluding this area. The counts in the jackknife sub-sample $TT_k$ are related to the overall number of counts in the full survey $TT_{\rm tot}$ and the above quantities by
\begin{equation}
    TT_k = TT_{\rm tot} - AA_k - CC_k.
    \label{eq:ttk}
\end{equation}
where in defining each of these pair counts we count each unique pair only once. The total number of auto- and cross-pairs can be related to their means over the jackknife samples by
\begin{equation}
    AA^{\rm tot} = N_{\rm jk} \overline{AA}
\end{equation}
and, as we account for double counting with the cross-pairs only while looking at the full sample, we need to divide the obtained estimate by 2 to be consistent with the auto-pairs:
\begin{equation}
    CC^{\rm tot} = \frac{N_{\rm jk}}{2} \overline{CC},
\end{equation}
where $\overline{AA} = \frac{1}{N_{\rm jk}}\sum_{k=1}^{N_{\rm jk}} AA_k$ and $\overline{CC} = \frac{2}{N_{\rm jk}}\sum_{k=1}^{N_{\rm jk}} CC_k$.

Following \citep{Mohammad_Percival}, we choose to define an estimator of the normalised auto-pairs $\theta_{{\rm a},k}$ in a specific realisation, such that $\overline{\theta}_{\rm a}=\overline{AA}$ by
\begin{equation}
    \theta_{{\rm a},k} = \frac{1}{N_{\rm jk}-1}\left(N_{\rm jk} \overline{AA} - AA_{k} \right)
\end{equation}
and the estimator of the normalised cross-pairs $\theta_{c,k}$ such that $\overline{\theta}_{\rm c}=\overline{CC}$ by
\begin{equation}
    \theta_{c,k} = \frac{2}{N_{\rm jk}-2} \left(\frac{N_{\rm jk}}{2} \overline{CC} - CC_k \right),
    \label{step1_CC}
\end{equation}
where it was taken into account that the cross-pairs contribute to the total estimate twice, while the auto-pairs only once.

We can then further compute for each jackknife realization the deviation from the mean value of the auto paircounts
\begin{equation}
\theta_{a,k}-\overline{\theta_{a}} = \frac{1}{N_{\rm jk}-1}\left(\overline{AA} - AA_{k}  \right)
\label{eq:auto_diff}
\end{equation}
and cross paircounts
\begin{equation}
\theta_{c,k}-\overline{\theta_{c}} = \frac{2}{N_{\rm jk}-2}\left(\overline{CC} - CC_{k}  \right).
\label{eq:cross_diff}
\end{equation}

We can now express how the covariance of each type of pair count can be represented in terms of the estimators above, if we assume the following definition for the covariance, where $DD_t$ are just some pair counts of type $t$:

\begin{equation}
\begin{split}
    \cov(DD_1,DD_2) = \sqrt{\frac{\overline{DD}_1\overline{DD}_2}{DD^{\rm tot}_{1}DD^{\rm tot}_{2}}}\, \frac{1}{N_{\rm jk}-1} \times\\ \times\sum_{k=1}^{N_{\rm jk}}(DD_{1k}-\overline{DD}_1)(DD_{2k}-\overline{DD}_2)
\end{split}
\label{eq:gen_jk_cov}
\end{equation}

By replacing $(DD_1, DD_2)$ by $(AA, AA)$ or $(CC, CC)$ or $(CC, AA)$ in Eq.~\ref{eq:gen_jk_cov} and using Eqs.~\ref{eq:auto_diff} and~\ref{eq:cross_diff}, one obtains:

\begin{equation}
    \cov(AA,AA) = \frac{N_{\rm jk}-1}{N_{\rm jk}} \sum^{N_{\rm jk}}_{k=1}\left( \theta_{{\rm a},k} - \bar{\theta}_{\rm a} \right)^2
    \label{eq:covAA_AA}
\end{equation}

\begin{equation}
    \cov(CC,CC) = \frac{(N_{\rm jk}-2)^2}{2N_{\rm jk}(N_{\rm jk}-1)} \sum^{N_{\rm jk}}_{k=1}\left( \theta_{{\rm c},k} - \bar{\theta}_{\rm c} \right)^2
    \label{eq:covCC_CC}
\end{equation}

\begin{equation}
    \cov(CC,AA) = \frac{(N_{\rm jk}-2)}{\sqrt{2}N_{\rm jk}} \sum^{N_{\rm jk}}_{k=1}\left( \theta_{{\rm c},k} - \bar{\theta}_{\rm c} \right)\left( \theta_{{\rm a},k} - \bar{\theta}_{\rm a} \right)
    \label{eq:covCC_AA}
\end{equation}

This gives all the components needed to compute the covariance of $TT$, using its definition in Eq.~\ref{eq:ttk}:

\begin{equation}
\begin{split}
    \cov(TT,TT) =  \cov(AA,AA)+\cov(CC,CC)+2 \cov(AA,CC) \\ = \frac{N_{\rm jk}-1}{N_{\rm jk}} \sum^{N_{\rm jk}}_{k=1}\left( \theta_{{\rm a},k} - \bar{\theta}_{\rm a} \right)^2  + \frac{(N_{\rm jk}-2)^2}{2N_{\rm jk}(N_{\rm jk}-1)} \sum^{N_{\rm jk}}_{k=1}\left( \theta_{{\rm c},k} - \bar{\theta}_{\rm c} \right)^2 + \\ + \frac{\sqrt{2}(N_{\rm jk}-2)}{N_{\rm jk}} \sum^{N_{\rm jk}}_{k=1}\left( \theta_{{\rm c},k} - \bar{\theta}_{\rm c} \right)\left( \theta_{{\rm a},k} - \bar{\theta}_{\rm a} \right)
\end{split}
\label{eq:covTT_TT}
\end{equation}

Note how the terms scale differently with the number of the jackknife regions. \cite{Mohammad_Percival} argue that this inconsistent scaling is the source of the bias that arises with the standard jackknife approach. In the next sections, we will see how adjusting this scaling can enable one to recover an unbiased covariance estimator and demonstrate the need for going beyond the Mohammad-Percival correction to get unbiased covariance estimators in all regimes of galaxy number density.

\subsubsection{Mohammad-Percival correction}
\label{subsec:MP}

\cite{Mohammad_Percival} proposed to weight the cross-pairs $CC$ in order to fix the mismatch in the scaling, as seen in Eq.~\ref{eq:covTT_TT}. With this  weight $\alpha$ multiplying all the $CC$ pair counts, the expression for $TT_k$ becomes
\begin{equation}
    TT_k = TT_{\rm tot} - AA_k - \alpha CC_k.
    \label{eq:ttk2}
\end{equation}

The definition of $\theta_{c,k}$ is then generalised to:

\begin{equation}
    \theta_{c,k}(\alpha) = \frac{2\alpha}{N-2\alpha} \left( \frac{N}{2} \overline{CC} - \alpha CC_{k}\right),
\end{equation}
which also changes slightly the mean of this quantity as $\overline{\theta}_{c}(\alpha)= \alpha \overline{CC}$.

Following the steps from equations~\eqref{step1_CC}, \eqref{eq:cross_diff} and~\eqref{eq:covCC_CC}, the modified expression for the covariance of the $CC$ paircounts weighted by $\alpha$ is 
\begin{equation}
    \cov(\alpha CC,\alpha CC) = \frac{(N_{\rm jk}-2\alpha)^2}{2 \alpha^2 N_{\rm jk} (N_{\rm jk}-1)} \sum^{N_{\rm jk}}_{k=1}\left( \theta_{c,k} - \bar{\theta}_{c} \right)^2
\end{equation}

We see that for $\alpha = 1$ we recover the ordinary jackknife, as it will remove the cross-pairs in the same way as it removes the auto-pairs. Alternatively, by choosing $\alpha = N_{\rm jk} / \left[2+\sqrt{2}(N_{\rm jk}-1)\right]$ we can achieve equal scaling for the first two terms. Therefore, under the assumption of $\cov(CC,AA) = 0$ we indeed have all the terms scaling with $N_{\rm jk}$ in same manner, which can be seen by rewriting the expression for $\cov(TT(\alpha),TT(\alpha))$ as
\begin{equation}
\begin{split}
    \\ \cov(TT(\alpha),TT(\alpha)) =  \cov(AA,AA)+\cov(\alpha CC,\alpha CC) \\ = \frac{N_{\rm jk}-1}{N_{\rm jk}} \sum^{N_{\rm jk}}_{k=1}\left( \theta_{{\rm a},k} - \bar{\theta}_{\rm a} \right)^2  +\\+  \frac{(N_{\rm jk}-2\alpha)^2}{2 \alpha^2 N_{\rm jk} (N_{\rm jk}-1)} \sum^{N_{\rm jk}}_{k=1}\left( \theta_{c,k} - \bar{\theta}_{c} \right)^2 
\end{split}
\end{equation}

\begin{figure}
\includegraphics[width=8cm]{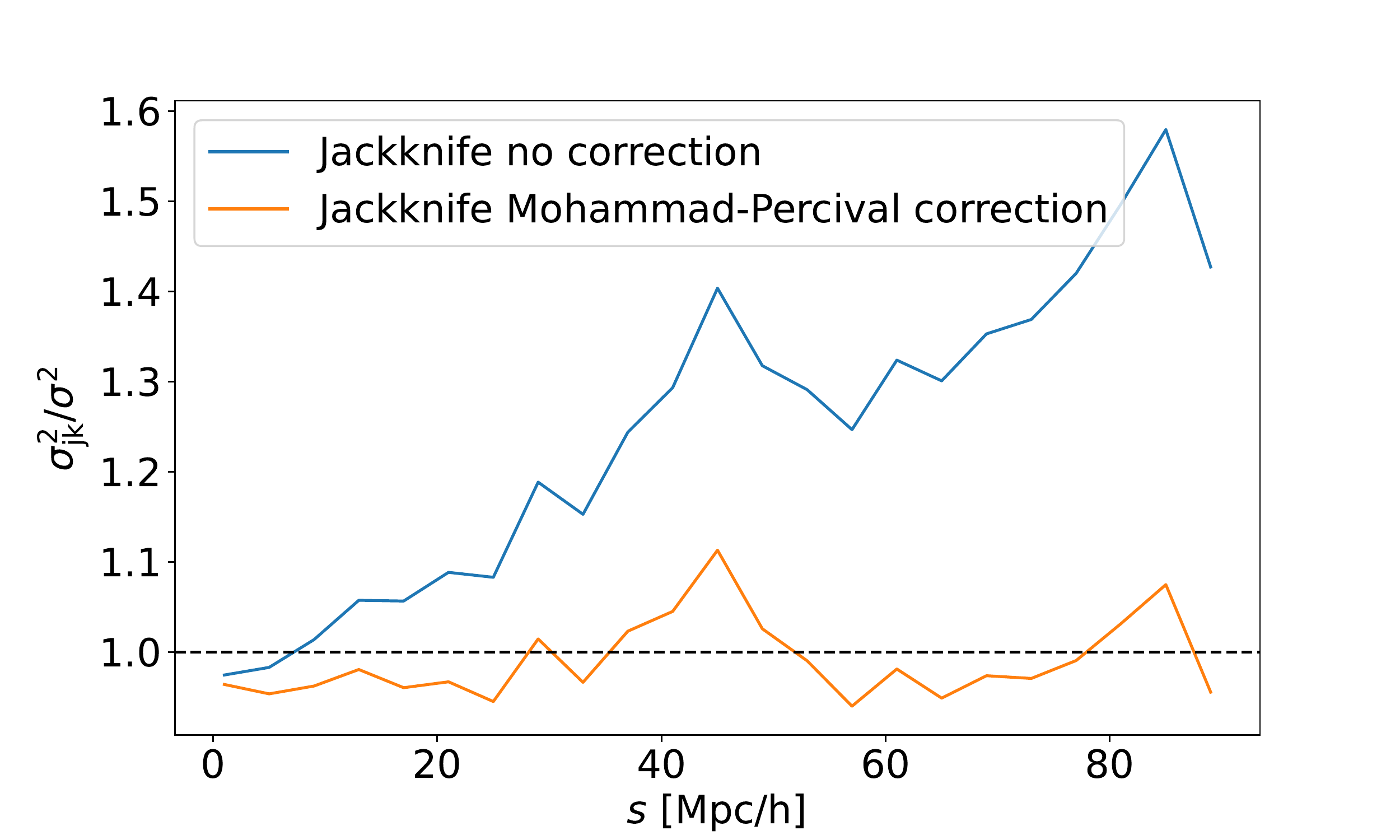}
\caption{Comparison of the accuracy in the estimate  of the diagonal elements of the covariance matrix for the real-space correlation functions as a function of scale obtained from 1000 cubic box independent mock catalogues. The ratio is the mean of the  diagonal elements obtained using different jackknife approaches to those obtained directly from the ensemble of mocks. The noticeable scale-dependent bias that is visible for the standard jackknife estimate is absent when the Mohammad-Percival correction is employed.}
\label{fig:rnd_jk}
\end{figure}

In order to illustrate the effect of introducing the $\alpha$ weighting of \cite{Mohammad_Percival}, we create 1000 Poisson random catalogues in a box with a size of 
1~Gpc/h, divide them into 125 cubic regions and then compute the covariance matrices of the real-space correlation function.
We do this for both the standard jackknife and jackknife with the Mohammad-Percival correction. The results are presented in Fig.~\ref{fig:rnd_jk}. We show the ratio of the mean of the diagonal elements, $\sigma^2~\equiv~C_{ii}$, of the covariance matrix between jackknife-based $\sigma_{\rm jk}$ and mock-based $\sigma$ (estimated directly using Eq.~\ref{eqn:cov_eq}), where the blue curve uses the standard jackknife and the orange one 
 includes the Mohammad-Percival correction. The standard jackknife is over-estimating the covariance with respect to that from the
mocks, while introducing the $\alpha$ weighting of \cite{Mohammad_Percival} for the cross-pairs removes this bias.

\subsection{Hybrid approach}
\label{subsec:fit_jk}
The real galaxy density has physical correlations and so galaxy distributions are not Poisson distributions. Therefore, the assumption of $\cov(CC,AA) = 0 $ is not valid. With the $\alpha$ weighting of the cross-pairs that was introduced in Section~\ref{subsec:MP}, Eq.~\ref{eq:covCC_AA} becomes
\begin{equation}
    \cov(\alpha CC,AA) = \frac{(N_{jk}-2\alpha)}{\sqrt{2}\alpha^2 N_{jk}} \sum^{N_{jk}}_{k=1}\left( \theta_{c,k} - \bar{\theta}_{c} \right)\left( \theta_{a,k} - \bar{\theta}_{a} \right).
\end{equation}

\begin{figure*}
    \centering
    \includegraphics[width=1.\linewidth]{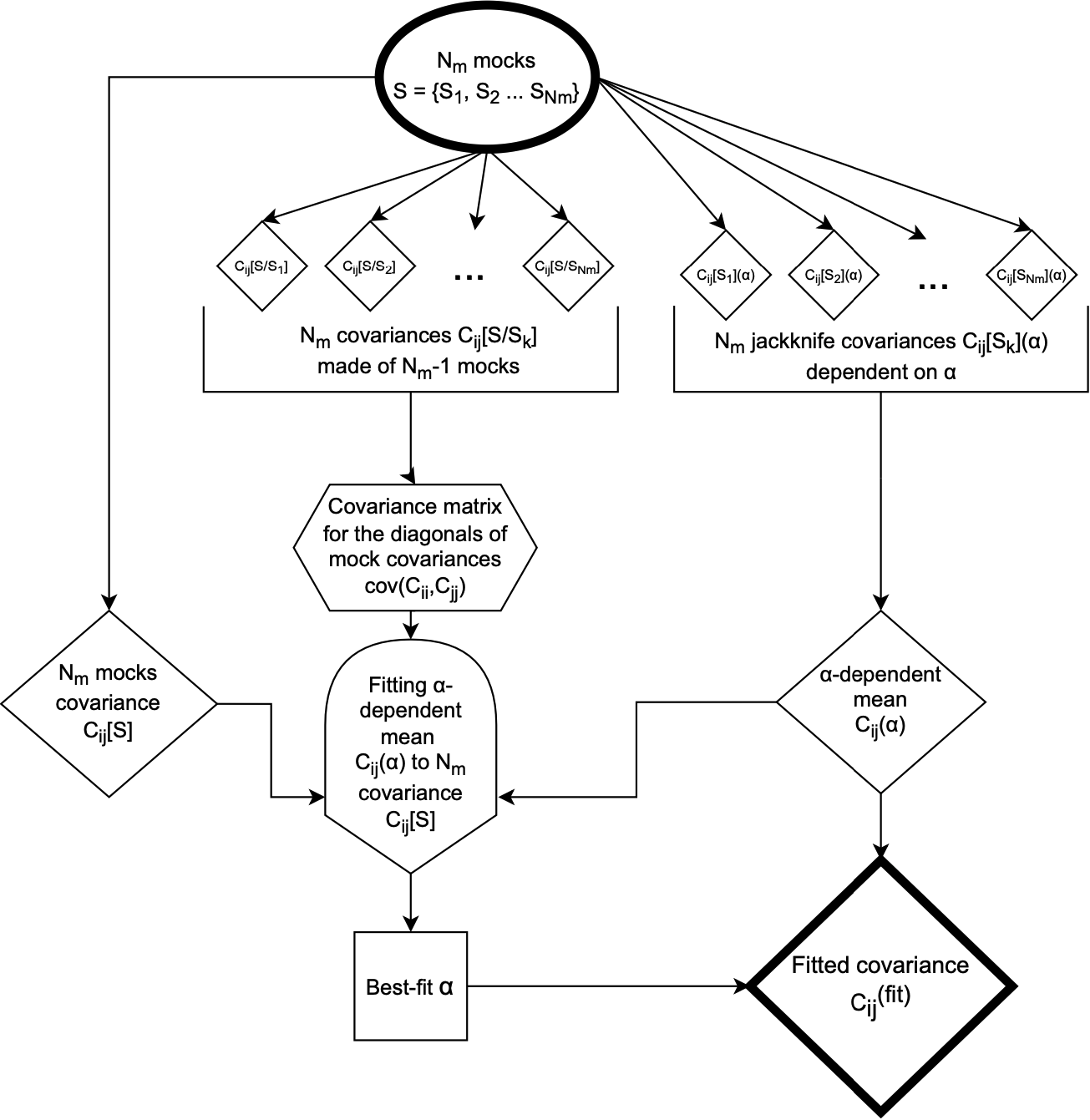}
    \caption{Schematic describing the procedure to obtain the fitted covariance $C^{ij}_{\rm fit}$ as defined in Eq.~\eqref{eqn:fit_eq} and discussed in Section~\ref{subsec:fit_jk}.}
    \label{fig:FitCovScheme}
\end{figure*}

We can see that adopting any general fixed value of $\alpha$ unfortunately leaves the scaling of $\cov(CC,AA)$ different from those of $\cov(AA,AA)$ and $\cov(CC,CC)$, so, in order to try to recover the benefits of the Mohammad-Percival approach, we are treating $\alpha$ as a free parameter.
We propose therefore to augment the  jackknife method with 
$\alpha$ weighting where the value of $\alpha$ is
tuned by fitting the covariance estimate from a limited number of mocks. A scheme that represents the approach is shown in Fig.~\ref{fig:FitCovScheme}. First, let us assume we have a set of $N_m$ mocks $S = \{ S_1... S_{N_m} \}$. Then, $S/S_k$ denotes the set of mocks with the $k^{\rm th}$ mock removed. Then, we refer to the mock covariance from such a set $S/S_k$ as $C_{ij}[S/S_k]$. We also introduce the $\alpha$-dependent jackknife  covariance obtained from a mock $S_k$ with a chosen $\alpha$ weighting as $C_{ij}[S_k](\alpha)$, from correlation functions constructed with counts following eq.~\eqref{eq:ttk2}.

Having that in our possession, we are able to estimate the uncertainty on the diagonal elements of the covariance $\Xi_{ij}({\diag}(C))$. First, we resample the given set of mocks and produce $N_m$ covariances $C_{ij}[S/S_k]$. Then we compute the covariance matrix of the diagonals $\Xi_{ij}({\diag}(C))$, where we limit ourselves to the diagonal elements as there are not enough degrees of freedom to build a covariance of matrices \citep{Wishart1928}:

\begin{multline}
\Xi_{ij}(\diag(C)) = \cov(C_{ii},C_{jj}) = \\ = \frac{N_{\rm m}-1}{N_{\rm m}}\sum^{N_{\rm m}}_{k=1} (C_{ii} [S/S_k] - C_{ii} [S])(C_{jj} [S/S_k] - C_{jj} [S])
\end{multline}

In general $N_{\rm m}$ should be greater than the number of elements in the fitted part of the covariance. However, in the case of a small $N_{\rm m}$, one can restrict this to just the diagonal elements of $\Xi_{ij}$, to ensure that covariance matrix stays non-singular.
The next step consists of finding which specific $\alpha$ is needed to obtain a realisation of the covariance matrix to describe $C_{ij}[S]$. First, we can write the $\alpha$ dependent estimator of the covariance $C_{ij}(\alpha)$ based on the mean of $N_{\rm m}$ $\alpha$ dependent jackknife covariances:
\begin{equation}
C_{ij}(\alpha)=\frac{1}{N_{\rm m}}\sum^{N_{\rm m}}_{k=1}C_{ij}[S_k](\alpha)   
\label{eqn:alpha_eq}
\end{equation}

Then, the $\chi^2$ of the $C_{ii}(\alpha)$ describing the $C_{ii}[S]$ can be written as:

\begin{equation}
    \chi^2_{C} (\alpha) = \sum_{ij} (C_{ii}(\alpha)-C_{ii}[S])\left(\Xi^{-1}\right)_{ij}(C_{jj}(\alpha)-C_{jj}[S])
\end{equation}

Following that, we minimise $\chi^2_C$ by varying $\alpha$, such that we obtain $\chi^2_{C}(\alpha_{\rm min}) = {\rm min} (\chi^2_C(\alpha))$.
To justify using the Gaussian likelihood in this procedure, we first notice that we are using only the diagonals of the covariance matrix. That allows us, with sufficiently large $N_{\rm m}$, to approximate the distribution of the separate bins of the diagonals $C_{ii}$ with a Gaussian.

Therefore, our proposed estimator of the $\alpha$ dependent covariance matrix $C^{(\rm fit)}_{ij}$ can be defined as:

\begin{equation}
C^{(\rm fit)}_{ij} = C_{ij}(\alpha_{\rm min}) =\frac{1}{N_{\rm m}} \sum^{N_m}_{k=1} C_{ij}[S_k](\alpha_{\rm min})
\label{eqn:fit_eq}
\end{equation}
While only the diagonal of $C^{(\rm fit)}_{ij}$ are used when fitting for $\alpha$, all the elements of 
$C^{(\rm fit)}_{ij}$ are consistently adjusted with the value of $\alpha$ that is found.
In the original Mohammad-Percival approach, the contribution of the cross-pairs to the covariance is adjusted to match that of the auto-pairs. Our hybrid approach allows us to adjust the cross-pair contribution on the $\alpha$ weighted covariance so that the covariance matches the one obtained from the limited set of mocks. We will show in the next section that by doing so, we can greatly reduce the bias that can appear for dense samples when using the fixed $\alpha$ weighting of \cite{Mohammad_Percival}. However, the hybrid approach does require more than a single mock to create a covariance estimate, but in the next section we will also show that the number of mocks needed is significantly reduced compared to a purely mock-based approach.

\section{Tests on mocks}
\label{sec:test-mocks}

We test the performance of the fitted jackknife method with respect to other covariance matrix estimation methods on different sets of mocks that include RSD and some geometrical effects that we will describe in subsequent sections.
For each specific set of mocks we also generate a set of matching random synthetic catalogues.

\begin{figure}
    \centering
    \includegraphics[width=8cm]{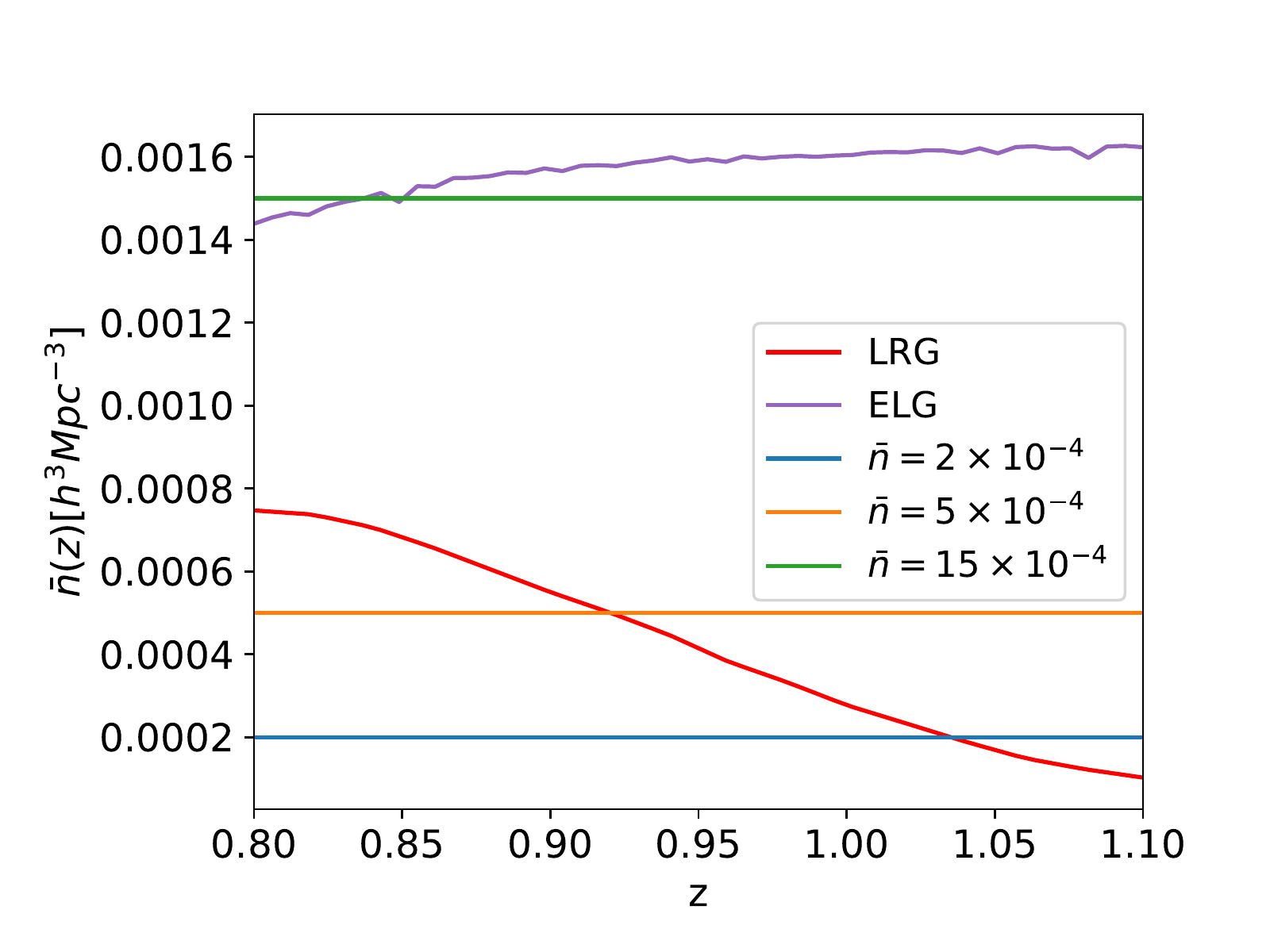}\vspace{-0.5 cm}
    \caption{Number density dependence on redshift for different datasets used. The lognormal mock samples were chosen to have a constant density selection function, to simplify the matters, while LRG and ELG mock samples follow the expected values from the corresponding DESI survey subsets.}
    \label{fig:nbars}
\end{figure}

In section~\ref{subsec:meth_tst} we present the methodology of the tests that we perform on our mocks. In section~\ref{subsec:log_tst}, a set of tests is performed on lognormal mocks produced by the \texttt{MockFactory} code\footnote{\url{https://github.com/cosmodesi/mockfactory}} with three number densities to explore shot noise-dominated and sample variance-dominated regimes, but also to mimic the DESI LRG and ELG samples.
In section~\ref{subsec:ez_tst} approximate EZmocks mimicking the DESI LRG and ELG samples are used to provide a mock-based covariance matrix which has the level of statistical precision of expected from the DESI Year-5 data. The corresponding number densities can be seen in Fig.~\ref{fig:nbars} for LRG EZmocks in red, ELG EZmocks in purple and the different lognormal mocks at $\bar{n}=(2,~5,~15)~\times~10^{-4}$~[Mpc/h]$^{-3}$ in blue, orange and green respectively. We use 1500 lognormal mocks for each space density, and 1000 ELG and LRG EZ mocks respectively.

\subsection{Methodology}
\label{subsec:meth_tst}
Both the random and data samples are divided into $N_{\rm jk}=196$ jackknife regions (the results, shown in Sec.~\ref{subsec:log_tst}, are not sensitive to $N_{\rm jk}$) and FKP weights $w_i$ for each point $i$ in the dataset are assigned as follows:
\begin{equation}
    w_i = \frac{1}{1+\bar{n}_i P_0}
\end{equation}
where $P_0 = 10^4 h^3 {\rm Mpc}^{-3}$ is the power spectrum estimate at the given redshift. The FKP weights~\citep{FKP} minimize the variance of the power spectrum estimate for samples that have a number density that varies with redshift.
Then, the correlation functions are computed using \texttt{pycorr}~\footnote{\url{https://github.com/cosmodesi/pycorr}} for both the samples and the jackknife realisations, which allows us to obtain 
$C_{ij}$, $C_{ij}(\alpha)$ and $C^{(\rm fit)}_{ij}$, defined in eq.~\eqref{eqn:cov_eq}, eq. \eqref{eqn:alpha_eq} and eq.~\eqref{eqn:fit_eq}.

%In order to test the robustness and precision of different covariance estimators, 

%We first create 30 fitted covariances. Then we randomly select 50 mocks.  These mocks are fitted with produced covariances in all possible combinations, allowing us to have 1500 different estimations of the cosmological parameters from different mocks obtained with different covariances.
%To have the same test for the jackknife covariances, we are randomly selecting 30 jackknife covariances and repeat the same procedure, obtaining the same total number of fits. 
%Unfortunately for the mock covariance, only 1 covariance matrix is available, so we fit all of the 1500 mocks, in order to obtain the same number of fits as for previous estimation techniques. The procedure is the same for the approximate mocks, with the only difference being that we only have 1000 mocks, so for the tests of fitted and jackknife covariance estimators we are using only 20 different covariances.

\begin{figure}
    \centering
    \includegraphics[width=1.\linewidth]{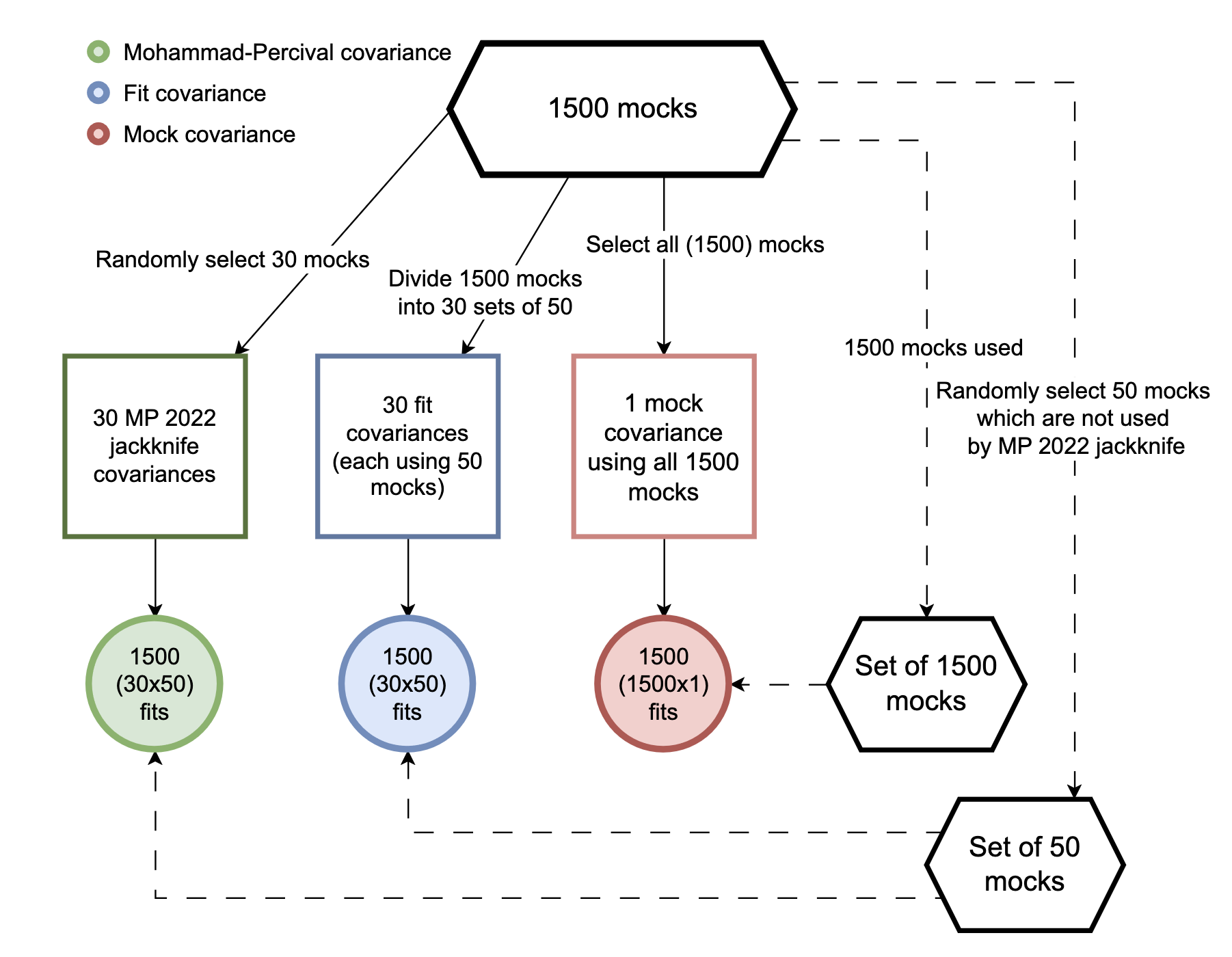}
    \caption{Schematic view of the procedure to test different covariance matrix estimators, as described in section~\ref{subsec:meth_tst}.}
    \label{fig:scheme_test}
\end{figure}

In order to test the robustness and precision of different covariance estimators using our set of lognormal mocks we have used the procedure described in Fig.~\ref{fig:scheme_test}. There we have created a set of 30 fitted and jackknife covariances and inferred cosmological parameters from 50 randomly selected mocks. In total we have 1500 pairs of covariances and mocks, which give us a set of 1500 fits for both the jackknife and fitted covariance approaches. As the mocks have the same cosmological parameters, and the covariances are considered estimators of the same underlying "true" covariance matrix, we then compared the spread of parameter values and their uncertainties to the one obtained from fitting separately each of the 1500 mocks to the conventional mock-based covariance matrix. 
The same is then repeated for the approximate mocks, with the difference that this time we have only 1000 mocks, bringing us to the sets of 20 fitted and jackknife covariances.

\subsection{Lognormal mocks}

\begin{figure}
    \centering
    \includegraphics[width=1\linewidth]{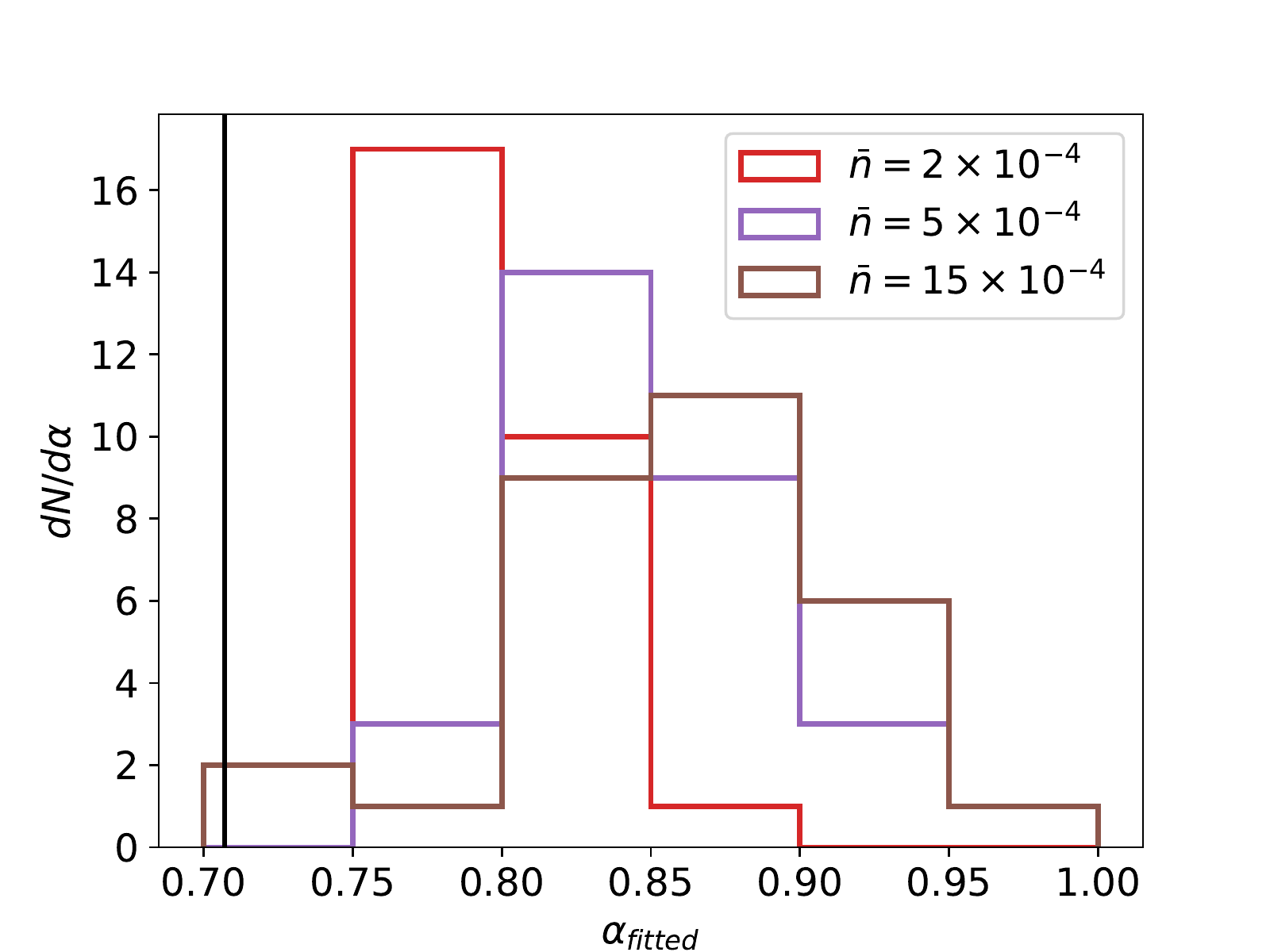}
    \caption{Histogram of the $\alpha$ parameter fitted from 50 mocks for lognormal mocks with $\overline{n} = 2\times 10^{-4}, 5\times 10^{-4} $and $15\times 10^{-4} \, h^3 {\rm Mpc}^{-3}$. The vertical black line shows the value of $\alpha = N_{\rm jk} /(2+\sqrt{2}(N_{\rm jk}-1))$}
    \label{fig:alpha_distr}
\end{figure}

\label{subsec:log_tst}
\begin{figure}
    \centering
    \includegraphics[width=1\linewidth]{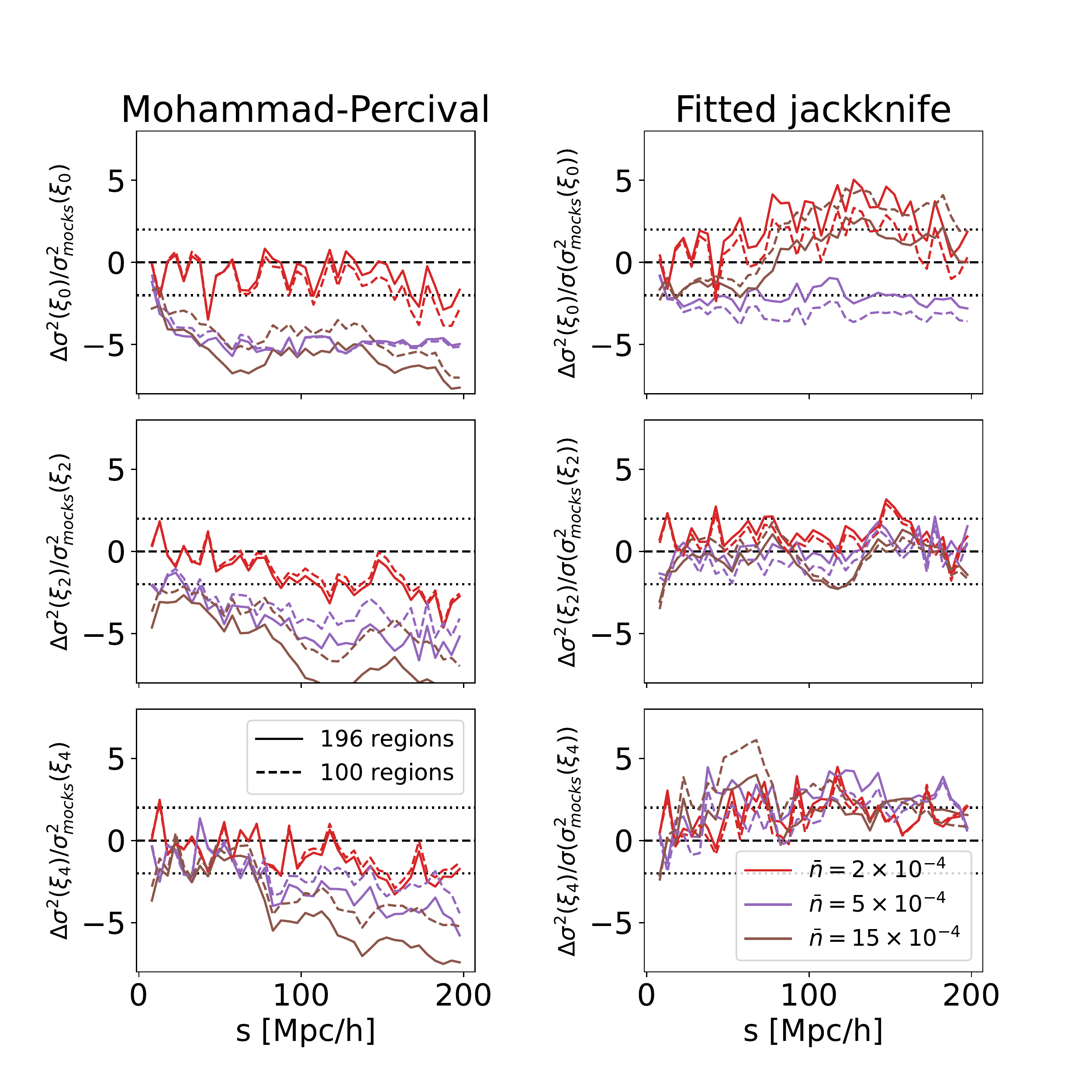}\vspace{-0.8 cm}
    \caption{The average of the quantity defined in Eq.~(\ref{eqn:def_bias})  representing the bias of the specific covariance estimation approach plotted as a function of separation, $s$, for various number densities. }
    \label{fig:covs_mp}
\end{figure}

In order to quickly test our approach with different parameters, such as number density, we produce a set of lognormal mocks which are often used as a simple approximation to the non-linear density field that evolves from Gaussian initial conditions.
The lognormal distributed density contrast $\delta (\vec{x})$ is related to a Gaussian field $G(\vec{x})= \ln[1+\delta(\vec{x})] - \langle \ln[1+\delta(\vec{x})]\rangle$ as:

\begin{equation}
    \delta(\vec{x}) = e^{-\langle G^2 \rangle + G(\vec{x})}-1
    \label{eq:lognormal_delta}
\end{equation}

The two-point correlation function $\xi(r)$ is related to the correlation function of the Gaussian field $\xi_G(r)$ as:

\begin{equation}
    \xi_G(r) = \ln[1+\xi(r)]
    \label{eq:lognormal_xi}
\end{equation}

So, a fiducial power spectrum $P(k)$ can be transformed into the correlation function $\xi(r)$, which is then converted to the correlation function of the Gaussian field using eq.~\eqref{eq:lognormal_xi}. We Fourier transform it to the power spectrum $P_G(k)$ and eventually generate the Fourier space Gaussian field $G(k)$ as:
\begin{equation}
    G(k) = \sqrt{\frac{P_G(k)V}{2}}(\theta_r+i\theta_i)
\end{equation}
where $\theta_r, \theta_i$ are Gaussian random variables with unit variance and zero mean, and $V$ is the volume of the simulation. After simulating the Fourier Gaussian field $G(k)$ on the grid, we then use Fast Fourier Transform (FFT) to transform it and obtain the regular configuration space Gaussian field $G(x)$. This is then transformed into the over-density field using eq.~\eqref{eq:lognormal_delta}. The expectation value for the number of galaxies in a particular cell is computed given a fixed mean number density $\bar{n}$, and galaxies are then drawn using the Poisson distribution and placed randomly the cell. Velocities are then assigned using the 
linearised continuity equation:
\begin{equation}
    a(t)\frac{\partial \delta(\vec{x})}{\partial t} + \vec{\nabla} \cdot \vec{v}(\vec{x}) = 0
\end{equation}
where $a(t)$ is a scale factor, and which is solved using Zeldovich approximation~\citep{Zeldovich}.

Eventually, the RSD effect is modelled at a chosen redshift using the velocity information by affecting the coordinates of the galaxy $x^i$ as:
\begin{equation}
    x^i_{\rm rsd} = x^i + f (\vec{n}.\vec{v})n^i
\end{equation}
where $x^i_{\rm rsd}$ are the redshift-distorted coordinates, $f$ is the linear growth rate of structure, $\vec{v}$ is the velocity of the galaxy, and $n^i$ is the line of sight.

\subsubsection{Dependence on number density}

We create 3 sets of lognormal mocks, each set containing 1500 realisations, for number densities $\bar{n} = 2\times10^{-4}$, $5\times10^{-4}$ and $ 15\times10^{-4} h^3{\rm Mpc}^{-3}$ at $z=1$. Each of the realisations is made from a cubic box with a volume of $(2 {\rm Gpc}/h)^3$ with grid of size $384^3$ and fiducial cosmology with $h = 0.674$, $\sigma_8 = 0.816$ and $\Omega_{\rm m}^{(0)} = 0.31$. The CLASS code~\citep{class} is used to generate the initial power spectrum. Redshift space distortions are then added, and each box is cut to have a footprint that covers $15 \%$ of the full sky. Each mock is then analysed in the redshift range from 0.8 to 1.2, and the corresponding randoms are generated, which are about 4 times denser than the data mocks.
The procedure to obtain the fitted jackknife covariance is summarised in Fig.~\ref{fig:FitCovScheme} and explained in the previous section. Here, we use $N_{\rm m} = 50$ mocks. We measure correlation functions from the mocks in bins of $5 h^{-1}$Mpc. Fig. \ref{fig:alpha_distr} presents the $\alpha$ parameter value distribution, obtained from the fits of the covariances.

Fig.~\ref{fig:covs_mp} shows a measure of the relative bias $\Delta \sigma^2(\xi_{\ell})/\sigma(\sigma^2_{\rm Mock})$ between a jackknife-based covariance matrix and the mock-based covariance as a function of pair separation $s$. For simplicity we only consider the diagonal elements of each covariance matrix estimate. This relative bias is defined as
\begin{equation}
    \frac{\Delta \sigma^2(\xi_{\ell})}{\sigma(\sigma^2_{\rm Mock})} = \frac{\sigma^2(\xi_{\ell}) - \sigma^2_{\rm Mock}(\xi_{\ell})}{\sigma(\sigma^2_{\rm Mock}(\xi_{\ell}))},
    \label{eqn:def_bias}
\end{equation}
where $\sigma(\xi_{\ell})$ is the variance on a given multipole $l$ obtained from the jackknife method, $\sigma_{\rm Mock}(\xi_{\ell})$ is the variance on the same multipole obtained from the 1500 lognormal mocks and $\sigma(\sigma^2_{\rm Mock})$ is the uncertainty on the mock-based error bar, determined by applying the classical jackknife delete-one mock estimator to the set of mocks from which the covariance is estimated.

The left panel of Fig.~\ref{fig:covs_mp} shows this relative bias of the jackknife method with the Mohammad-Percival correction  while the right panel shows the result for our fitted jackknife method. In both cases, the monopole, $\xi_{0}$, is displayed in the top panel, the quadrupole, $\xi_{2}$, in the middle and the hexadecapole, $\xi_{4}$, in the bottom. The coloured lines show different number densities and the solid lines are the baseline configuration of 196 jackknife regions while the dashed lines show the test of using 100 jackknife regions instead. As expected, the underestimation slightly worsens with the increase in the number of jackknife regions, as predicted by eq.~\eqref{eq:covCC_AA}.

However,  as the number density $\bar{n}$ increases, the underestimation of the jackknife method with the Mohammad-Percival correction becomes more and more significant, especially for $\bar{n} = 15\times 10^{-4}\,h^3{\rm Mpc}^{-3}$. This underestimation is not visible on the jackknife covariance matrix estimates produced from the random catalogues as shown in Fig.~\ref{fig:rnd_jk}. As explained in the previous section, the clustering of the data leads to higher covariance due to additional covariance coming from cross-correlations between $CC$ and $AA$ pair counts. 

Additionally, there is no strong dependence on the number density for the fitted jackknife method which makes it more robust whatever the density regime of the galaxy sample of interest. It should be noted that for low-density regimes optimal $\alpha$ seems to be closer to the default value of Mohammad-Percival approach, and its fitted estimation in our method introduces additional uncertainty, which makes our method more imprecise as $\overline{n}(z)$ decreases.

\subsubsection{Effect on the cosmological parameters}

To test the performance of different covariance estimation techniques we infer $f\sigma_8$, $\alpha_{\parallel}$ and $\alpha_{\perp}$ by fitting the theoretical predictions for the multipoles to the ones from the mocks using covariances from estimators reviewed earlier. The fit  is performed using a 5-parameter model, which is based on Lagrangian Perturbation Theory and includes the linear growth rate $f\sigma_8$, Alcock-Paczynski parameters \citep{AlcockPaczynski} $\alpha_{\parallel}$ and $\alpha_{\perp}$, first- and second-order biases $b_1$, $b_2$ and the effective Fingers Of God parameter (FOG) $\sigma_{\rm FOG}$. The theoretical power spectrum $P_\mathrm{FOG}$ is obtained using the \texttt{MomentumExpansion} module of the\texttt{velocileptors} package \citep[for more details, see][]{velocileptors1, velocileptors2}. The Fingers-Of-God effect is modelled following \cite{Taruya}, as
\begin{equation}
   P_\mathrm{FOG}(\boldsymbol{k}) = \frac{1}{1+(\boldsymbol{k}\cdot \hat{\boldsymbol{n}} \: \sigma_\mathrm{FOG})^2/2} P(\boldsymbol{k}), 
\end{equation}
where $P(\boldsymbol{k})$ is the power spectrum without the FOG effect obtained with \texttt{velocileptors}, $\sigma_\mathrm{FOG}$ is the one-dimensional velocity dispersion and $\hat{\boldsymbol{n}}$ is the LOS direction unit vector. 
The power spectrum $P_\mathrm{FOG}(\boldsymbol{k})$ is then transformed into the 2-point correlation function $\xi^{\rm th}(s,\mu)$ using a Fast-Fourier-Transform and from that we compute the theoretical correlation function multipoles $\xi^{th}_{\ell}(s,\mu)$.

\begin{figure}
    \centering
    \includegraphics[width=1.\linewidth]{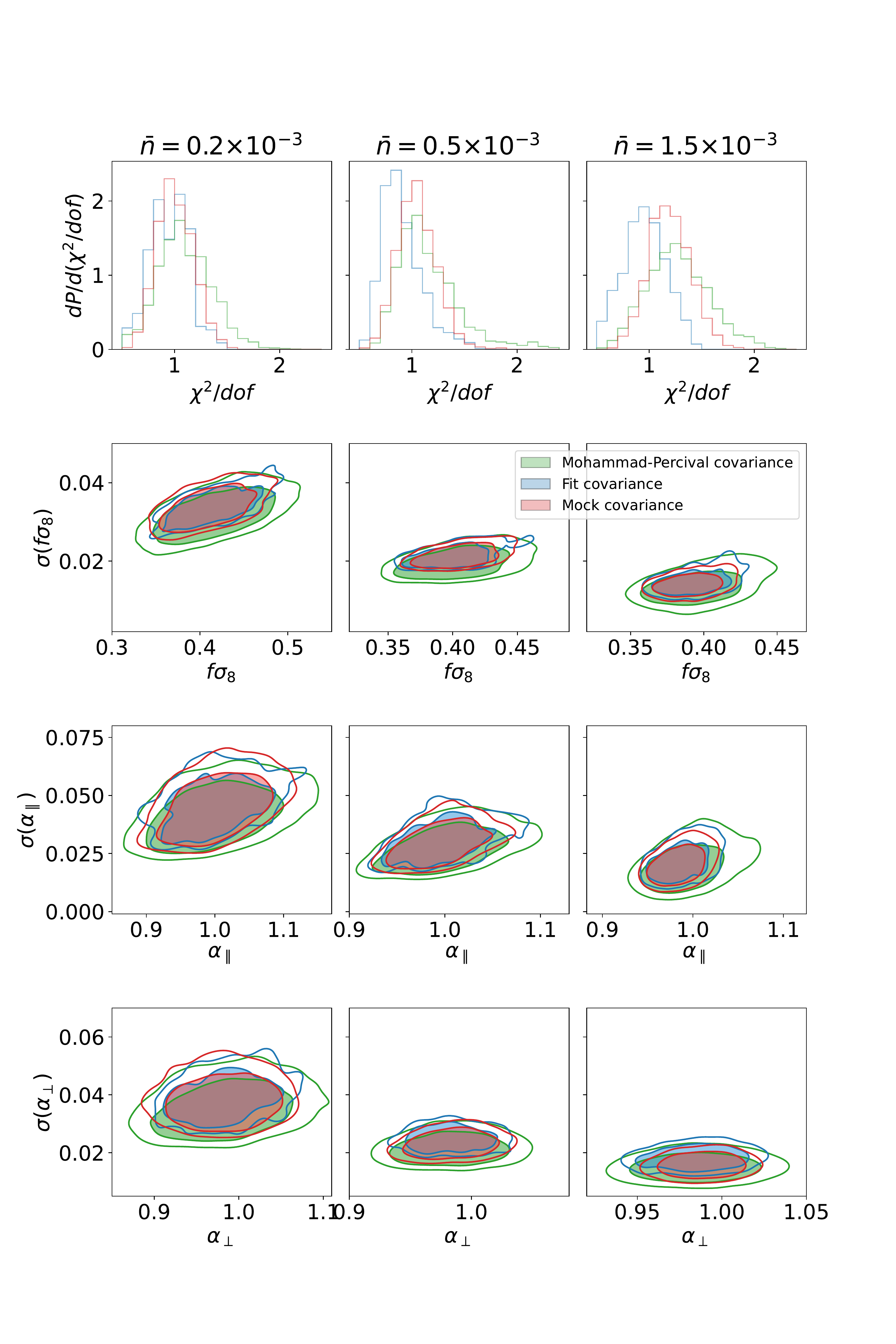} \vspace{-1.3 cm}
    \caption{The summary of the results from the cosmological fits from the lognormal mocks with varying density (one for each column and with density in (Mpc/h)$^3$ indicated at the top) for the three covariance matrix estimation methods: jackknife covariance with Mohammad-Percival correction in green, fitted jackknife covariance in blue and mock covariance in red. The top panels shows the histograms of the reduced $\chi^2$, while the three bottom ones show the marginalised 2D-distributions of parameters and their uncertainties for $f\sigma_8$, $\alpha_{\parallel}$ and $\alpha_{\perp}$, obtained from the set of fits described in the Section~\ref{subsec:meth_tst} }
    \label{fig:fits_spreads}
\end{figure}

\begin{figure}
    \centering
    \includegraphics[width=1.\linewidth]{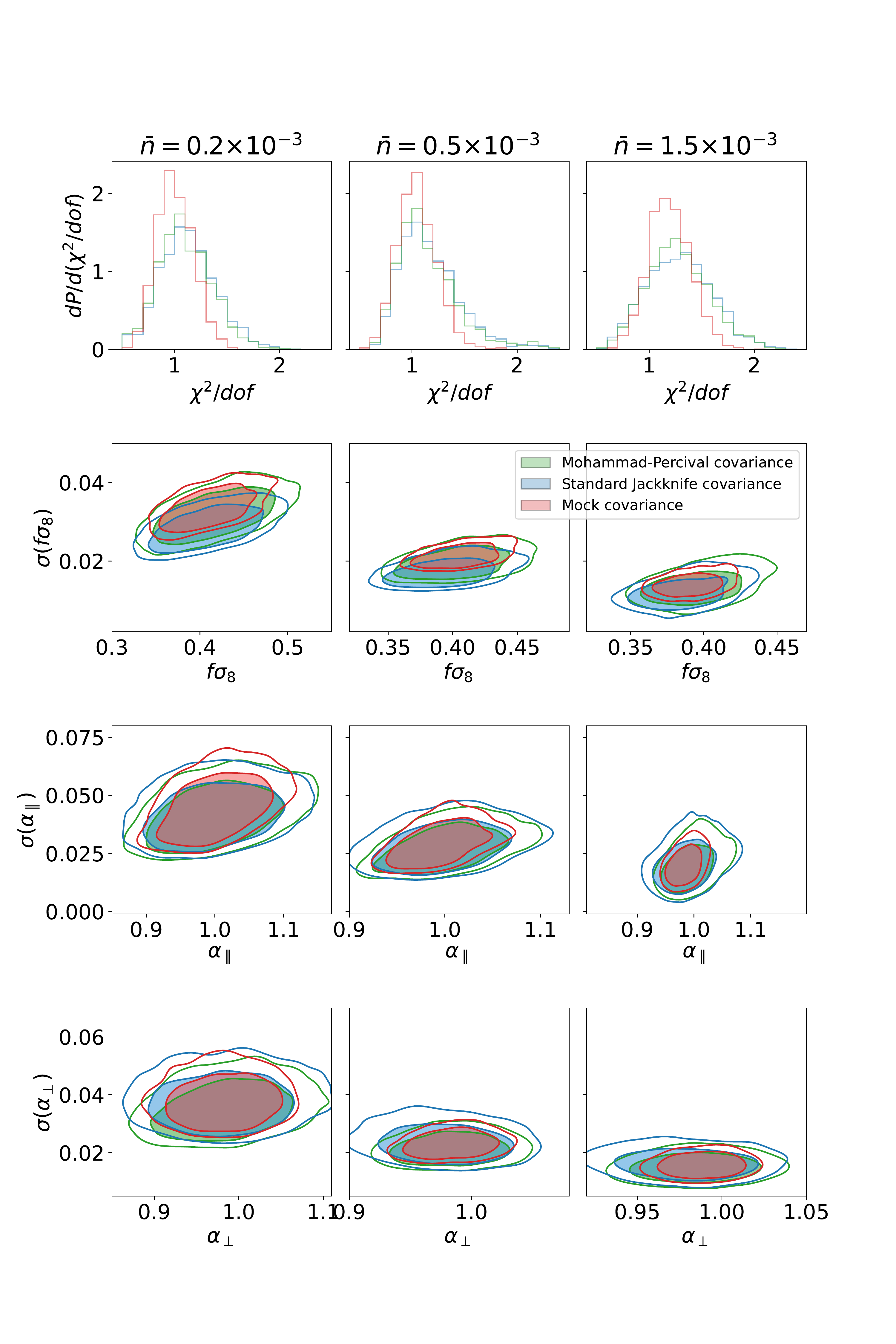} \vspace{-1.3 cm}
    \caption{The summary of the cosmological fits from the lognormal mocks with a varying density. Similar to Fig.~\ref{fig:fits_spreads} but for a slightly different set of covariance matrix methods: jackknife covariance with the Mohammad-Percival correction in green, mock covariance in red (the same contours as on Fig.~\ref{fig:fits_spreads}) and standard jackknife covariance in blue.}
    
    \label{fig:std_jk_spreads}
\end{figure}

\begin{table}
\begin{center}
\begin{tabular}{||c c c c ||}
\hline
$\bar{n}(z) (h^3 {\rm Mpc}^{-3})$ & Mock & Mohammad-Percival & Fit \\
\hline
\hline
$2\times 10^{-4}$ & 1.03 & 1.40 & 1.05 \\
\hline
$5\times 10^{-4}$& 0.99 & 1.42 & 1.05 \\
\hline
$15\times 10^{-4}$& 1.00 & 1.56 & 1.08 \\
\hline
\end{tabular}
\end{center}
\caption{For each of the estimation methods we tabulate the standard deviation $\sigma$ of $(f\sigma_{8i}-\overline{f\sigma_8})/\sigma_i(f\sigma_8)$, over independent fits, $i$. For the mock covariance method $\sigma \approx 1$ (as expected when all the fits are performed consistently with the same covariance), for the fitted covariance method it is also quite close to unity, but for the jackknife method $\sigma>1.4$, which shows a much higher degree of deviation from the truth.}
\label{tab:devs}
\end{table}

Once we have the correlation function multipoles $\xi_{\ell}(s_i)$, and covariance matrix $\Sigma^{\ell_1 \ell_2}_{ij} = C_{ij}$, $C^k_{ij}(\alpha) $ or $C^{(\rm fit)}_{ij}$, we can obtain the likelihood $L(p_1,...,p_n)$:

\begin{equation}
\begin{split}
    \log(L(p_1,..,p_n)) = \sum_{\ell_1,\ell_2} \sum_{i,j} \left(\xi_{\ell_1}(s_i)-\xi^{(\rm th)}_{\ell_1}(s_i)\right) \times \\ \times\left(\Sigma^{-1}\right)^{\ell_1 \ell_2}_{ij}\left(\xi_{\ell_2}(s_j)-\xi^{(\rm th)}_{\ell_2}(s_j)\right)
\end{split}
\end{equation}
where $\xi^{(th)}_{\ell}(s)$ is the theoretical prediction of the multipole for the set of $k$ parameters $\{a_1,..,a_k\}$ and we apply the Hartlap correction \citep{Hartlap} on the inverse of the covariance matrix such that the original uncorrected covariance matrix denoted as $C^{(\rm orig)}_{ij}$ and the corrected inverse covariance matrix $\Sigma^{-1}_{ij}$ are related by:
\begin{equation}
    \Sigma^{-1}_{ij} = \frac{n-p-2}{n-1}\left(C^{-1}\right)^{(\rm orig)}_{ij}
\end{equation}
where $n$ is the number of discrete samples, and $p$ is the number of entries in the data vector (number of bins used).
We use a likelihood maximisation method to find the $\chi^2$ minima using \texttt{iminuit} \citep{iminuit}. Errors are estimated from the region of $\Delta \chi^2 =1$ of the marginalized $\chi^2$ distribution, and they are allowed to be asymmetric. The choice of a frequentist method of analysis is motivated by its low computational cost.

In Fig.~\ref{fig:fits_spreads}, the first row shows the distributions of reduced $\chi^2$ for different choices of $\bar{n}$, and the other rows show the marginalised 2D-distributions of parameters and their uncertainties for respectively $f\sigma_8$, $\alpha_{\parallel}$ and $\alpha_{\perp}$.
The distributions of reduced $\chi^2$ show the goodness of the individual fits for the three methods. The contours in the bottom panel show how, for all the parameters, the spread from the Mohammad-Percival jackknife in green is in general much wider than the one from the mock covariance in red  both in terms of uncertainty and parameter values. While in case of the fitted jackknife covariance, the blue contours are very similar to the mock covariance ones. Presumably, this improvement comes from using 50 realisations rather than one. 
In Fig. \ref{fig:std_jk_spreads} we also show in the same form the performance from the standard jackknife in comparison with the Mohammad-Percival corrected jackknife and mock-based covariance. As expected, the standard jackknife produces slightly larger contours, which are noticeably shifted with respect to the mock covariance, especially for $f\sigma_8$. 
%On the plots of the distribution of the $\chi^2/{\rm dof}$ we can also notice that $\chi^2/{\rm dof}$ from fits with fitted jackknife covariance is slightly smaller than in other cases, and closer to 1. The difference grows with the $\overline{n}(z)$, but does not become significantly large, and the distributions are still compatible.

To additionally test the validity of our inference approaches, we will define the quantity
\begin{equation}
    x = \frac{\eta-\bar{\eta}}{\sigma(\eta)},
    \label{eq:pull}
\end{equation}
where $\eta$ is an inferred parameter from a specific fit, $\bar{\eta}$ is the mean from all the fits, and $\sigma(\eta)$ is the error estimation from a specific fit. 
The distribution of quantity $x$ is called a pull distribution. If $\eta$ follows a Gaussian distribution, the distribution of $x$ will form a normal distribution with $\bar{x}=0$ and $\sigma(x)=1$. 

For the mock covariance, we fit the 1500 available samples, while for the Mohammad-Percival jackknife and for the fitted jackknife 50 random mocks are fitted using 30 realisations of the covariance, under the assumption that all of the covariance estimators are probing the same underlying likelihood.

\begin{figure}
    \centering
    \includegraphics[width=1.\linewidth]{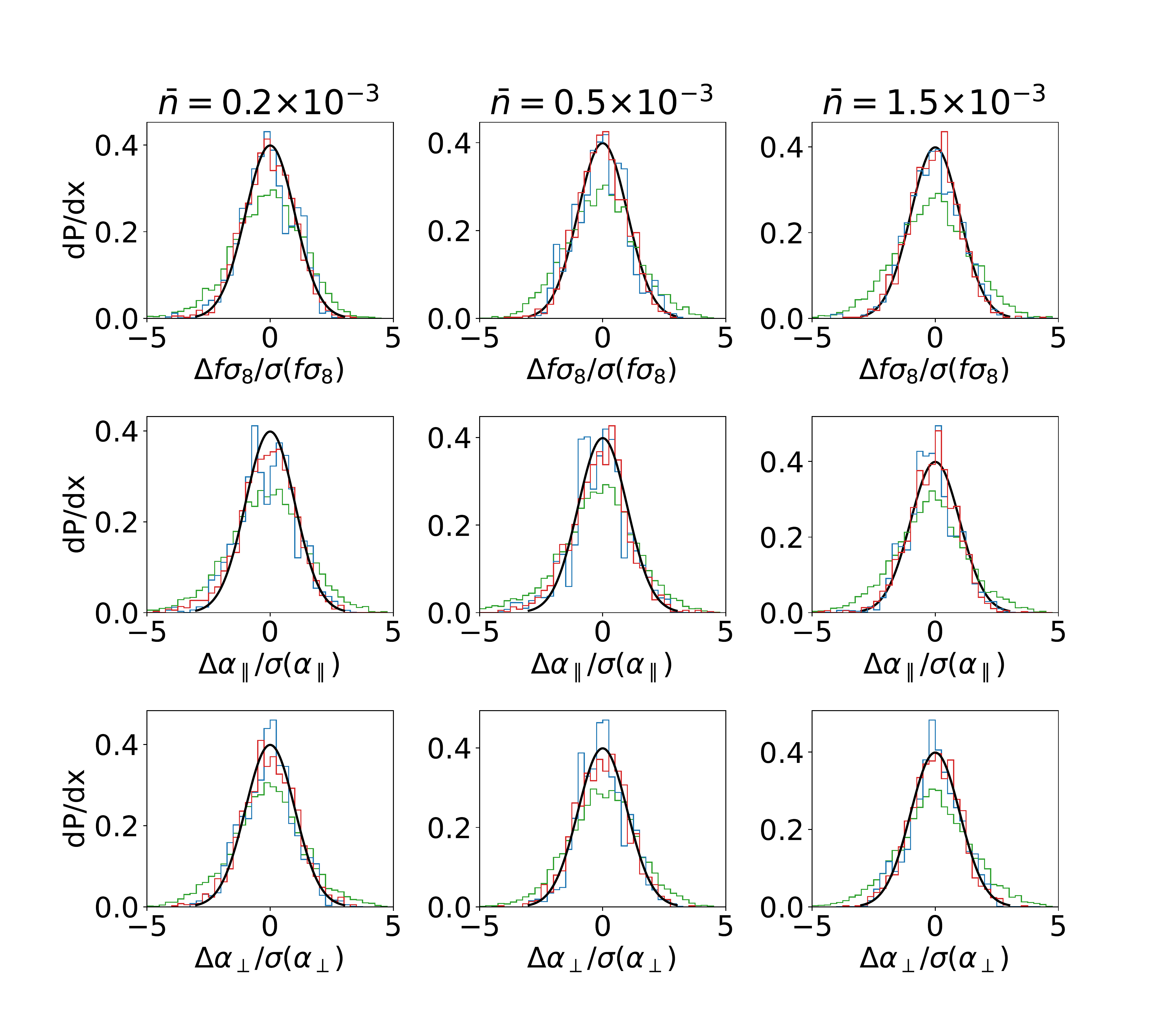} \vspace{-0.8 cm}
    \caption{Pull distributions for different covariance estimation techniques with results from fits on various lognormal mocks, shown for 3 different number densities indicated at the top in (Mpc/h)$^3$. Line colors follow those in Fig. \ref{fig:fits_spreads} }
    \label{fig:fits_hists}
\end{figure}

Pull distributions for $f\sigma_{8}$, $\alpha_{\parallel}$ and $\alpha_{\perp}$ are presented in Fig.~\ref{fig:fits_hists} for each number density of the lognormal mocks. The fitted jackknife and mock covariance pull distributions have Gaussian-shape with $\sigma = 1$ normal distributions as expected, while the pull distributions obtained when using Mohammad-Percival jackknife are slightly wider, which is due the covariance being less precise. We can see it quantitatively in the Table \ref{tab:devs}, where the standard deviations of the distributions from Fig.~\ref{fig:fits_hists} are presented.
That is due to various shifts of the distributions obtained from fitting to different jackknife covariances. This is not the case for the fitted approach, however.

Overall, for all the number densities, the performance of the fitted jackknife method using 50 mocks is much better than that of the standard jackknife with the Mohammad-Percival correction, and, most importantly, it gives similar performance as the covariance matrix created from 1500 mocks.

\begin{figure}
    \centering
    \includegraphics[width=1.\linewidth]{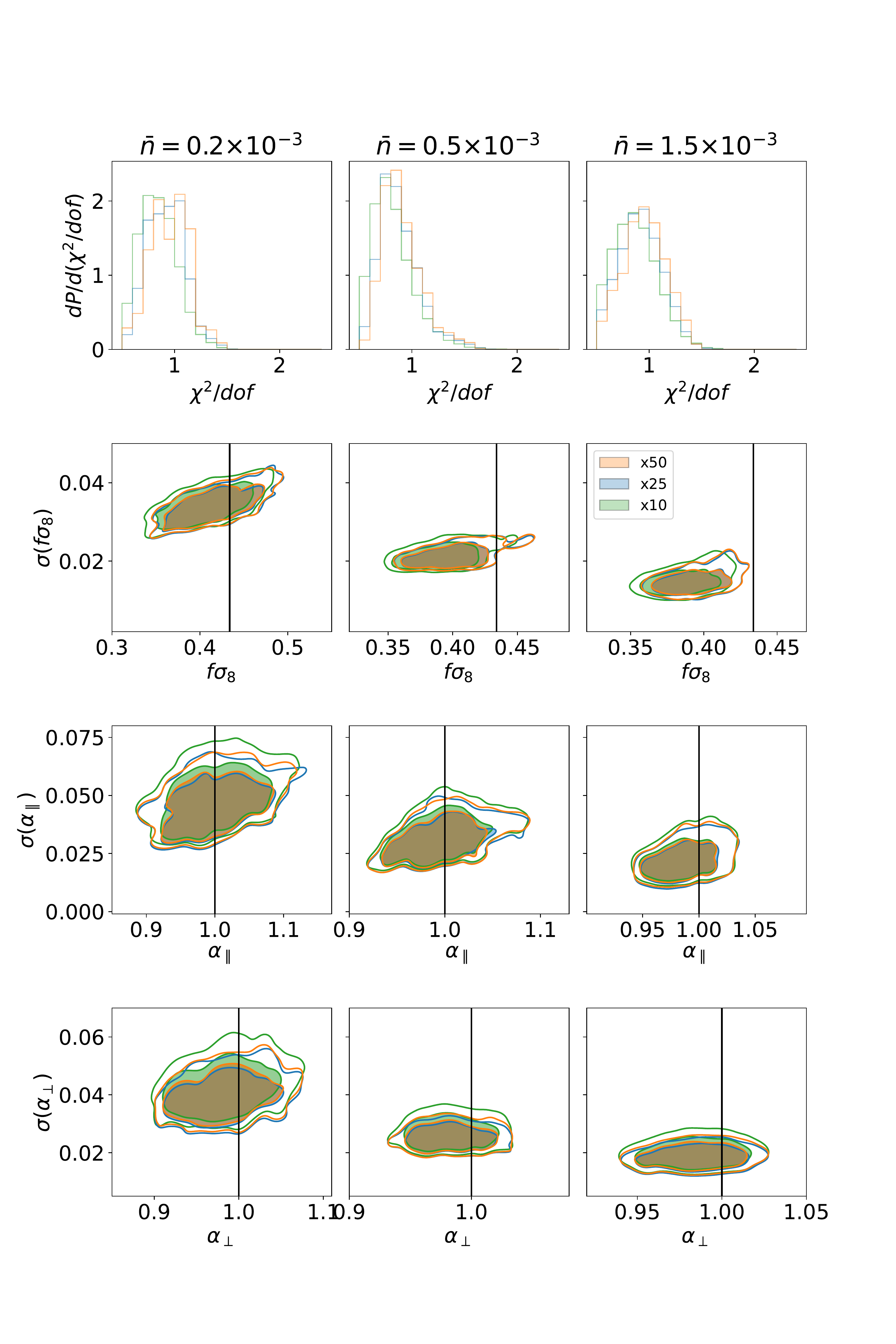} \vspace{-1 cm}
    \caption{The summary of the cosmological fits when using different numbers of mocks to obtain the fitted jackknife covariance: the default number of 50 mocks in red, 25 mocks in blue and 10 mocks in green. The figure is organised like Fig.~\ref{fig:fits_spreads}.}
    \label{fig:fits_spreads_vs}
\end{figure}

We also test the performance of the approach when varying the number of mocks used for producing the fitted covariance. We test using 10, 25 and 50 mocks and report the results on the cosmological fits in Fig.~\ref{fig:fits_spreads_vs}, following the same methodology as explained before for 50 mocks. The precision on the marginalised 2D contours of the cosmological parameters of interest starts to drop noticeably when 10 mocks are used, while it remains stable between 25 and 50 mocks.

\subsection{Approximate mocks}

\label{subsec:ez_tst}

Approximate mocks based on the extended Zeldovich approximation described in \cite{ez_mocks_eboss} are used to mimic the DESI LRG and ELG samples. These mocks are expected also to reproduce the clustering in the quasi-linear regime, although they are less accurate than N-body simulations. They provide a better representation of the real survey and better reproduce the non-Gaussian effects, which are not present in the lognormal mocks.
The EZmocks used here are built using a 4-parameter model that is calibrated to match the clustering of N-body simulations, the 25 AbacusSummit simulations designed to meet the DESI requirements~\citep{abacus}. The 4 model parameters are: (1) $\rho_{\rm c}$ - critical density required to overcome the background expansion; (2) $\rho_{\rm exp}$ - responsible for the exponential cut-off of the halo bias relation; (3) $b$ - argument in the power law probability distribution function $P(n) = const\times b^{n}$ of having $n$ galaxies in the limited volume; (4) $\nu$ is the standard deviation for the distribution modelling peculiar velocities.

\begin{figure}
    \centering
    \includegraphics[width=1.1\linewidth]{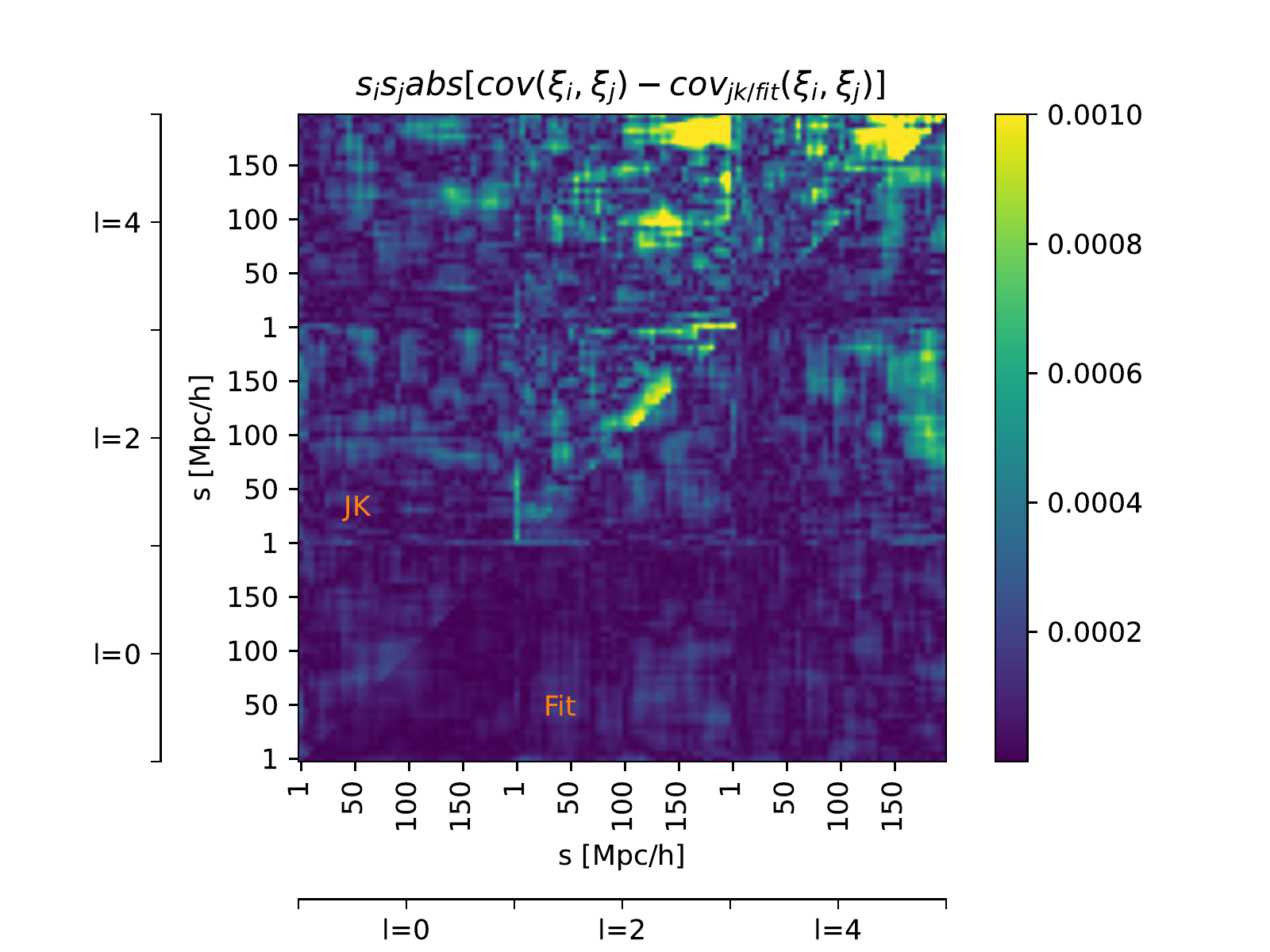}
    \caption{Comparison of the deviation of jackknife and fit covariances from the mock covariance  multiplied by a square of separation for multipoles $\ell=0,2,4$ for the EZ LRG mocks. }
    \label{fig:cov_comp_diff}
\end{figure}

\begin{figure}
    \centering
    \includegraphics[width=1.\linewidth]{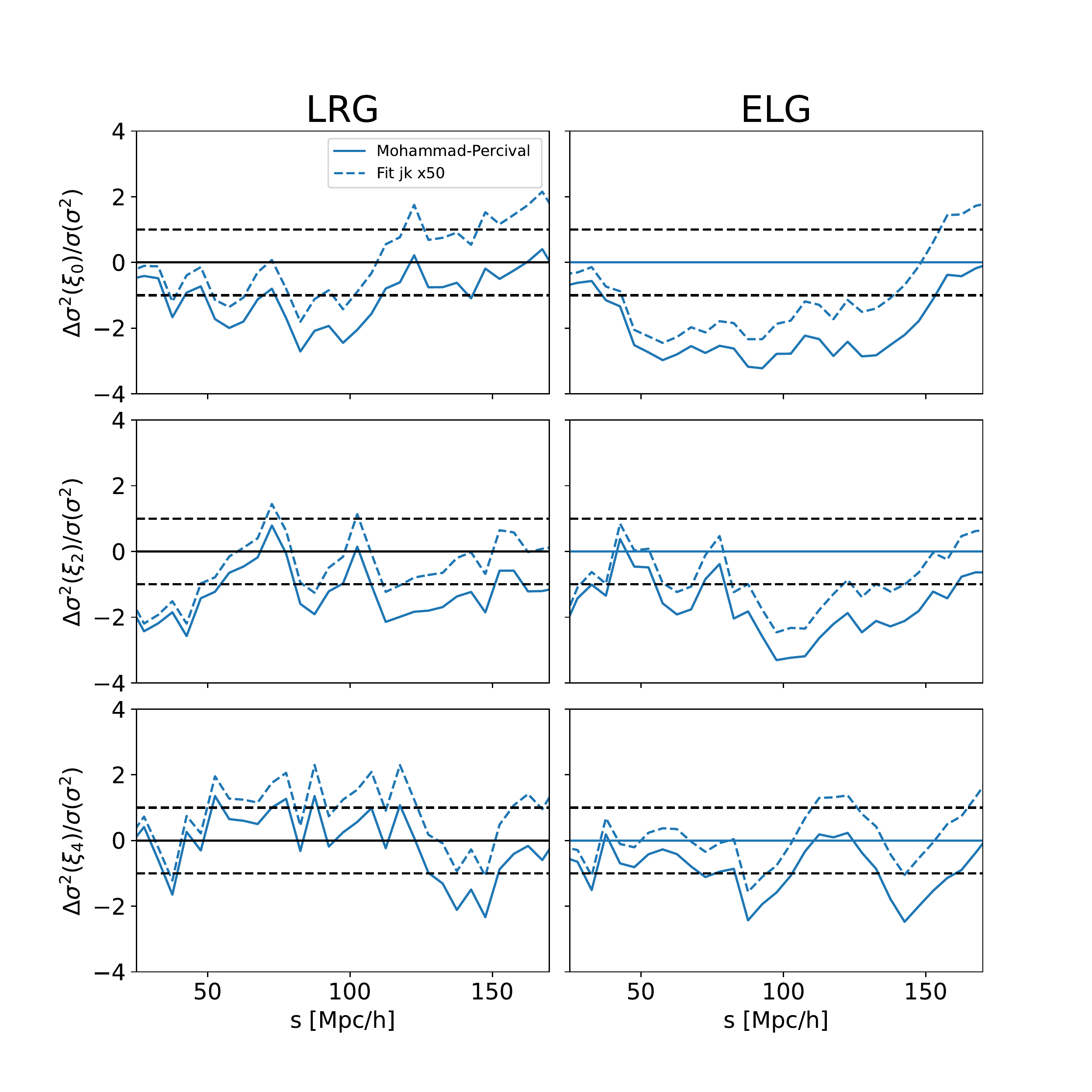}\vspace{-1 cm}
    \caption{The quantity defined in Eq.~(\ref{eqn:def_bias}) representing the bias of the specific covariance estimation approach plotted for three multipoles of LRG and ELG EZmocks (left and right panels respectively). Solid lines are with Mohammad-Percival correction and dashed lines for the fitted jackknife.}
    \label{fig:approx_diags}
\end{figure}

In this work, we use a set of 1000 EZmocks generated from N-body simulations with 6 Gpc/h box size. The fiducial cosmology employed is Planck 2018 \citep{planck2018}, and the boxes are generated at $z = 0.8$ for the LRGs and $z=1.1$ for the ELGs. We use the redshift range of $z=[0.8,1.1]$ and the mocks are cut to a footprint that reproduces that planned for the 5-year DESI data in order to match the expected final precision of the mock-based covariance matrix. The comparison of the difference with the mock covariance for the single realisation of the jackknife covariance and the fitted covariance is presented in Fig.~\ref{fig:cov_comp_diff}.

\begin{figure}
    \centering
    \includegraphics[width=1.\linewidth]{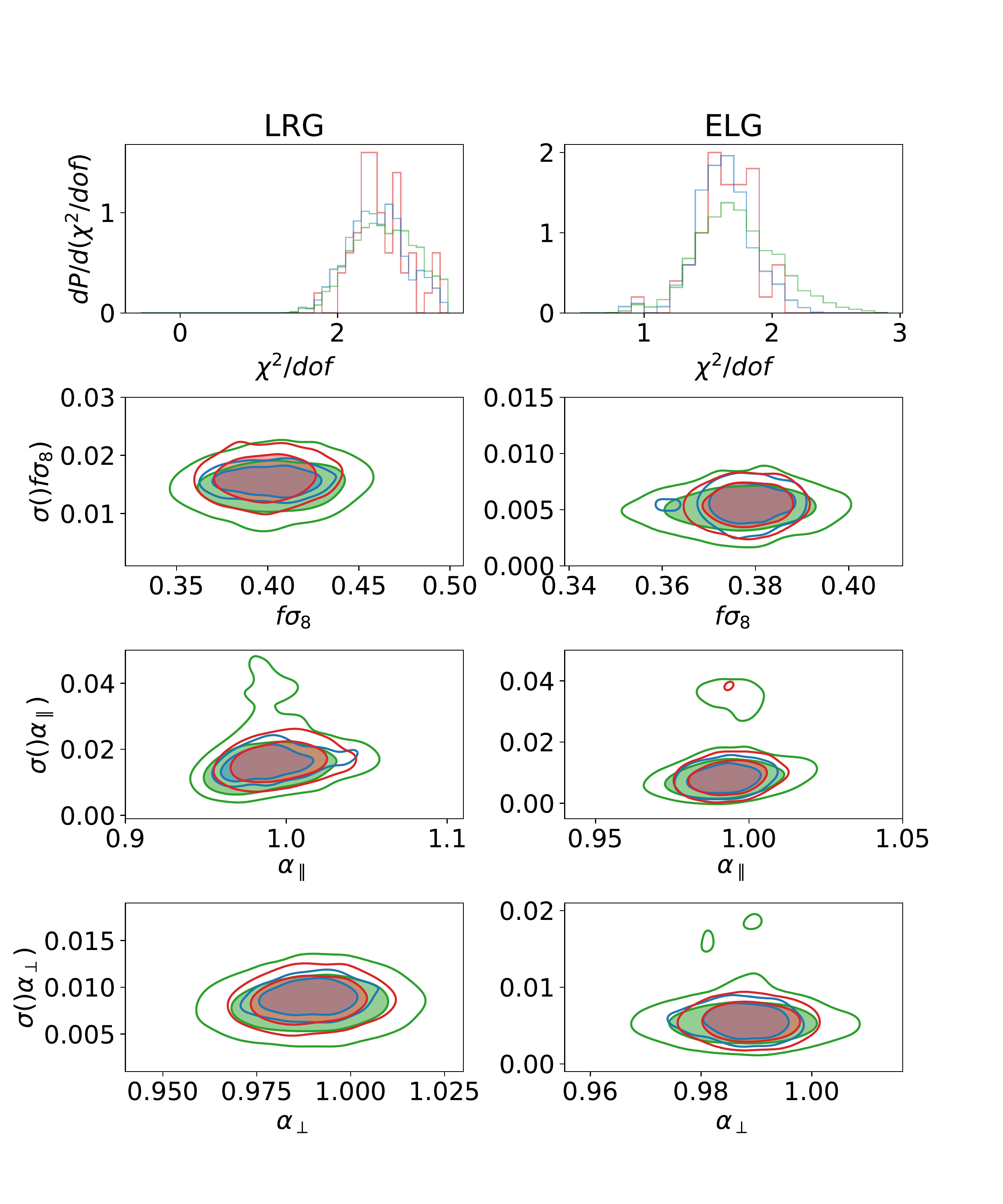} \vspace{-1 cm}
    \caption{The summary of the cosmological fits for the EZ mocks for LRGs and ELGs (left and right column respectively), similar to Fig.~\ref{fig:fits_spreads} layout.}
    \label{fig:approx_fits}
\end{figure}

\begin{figure}
    \centering
    \includegraphics[width=1.\linewidth]{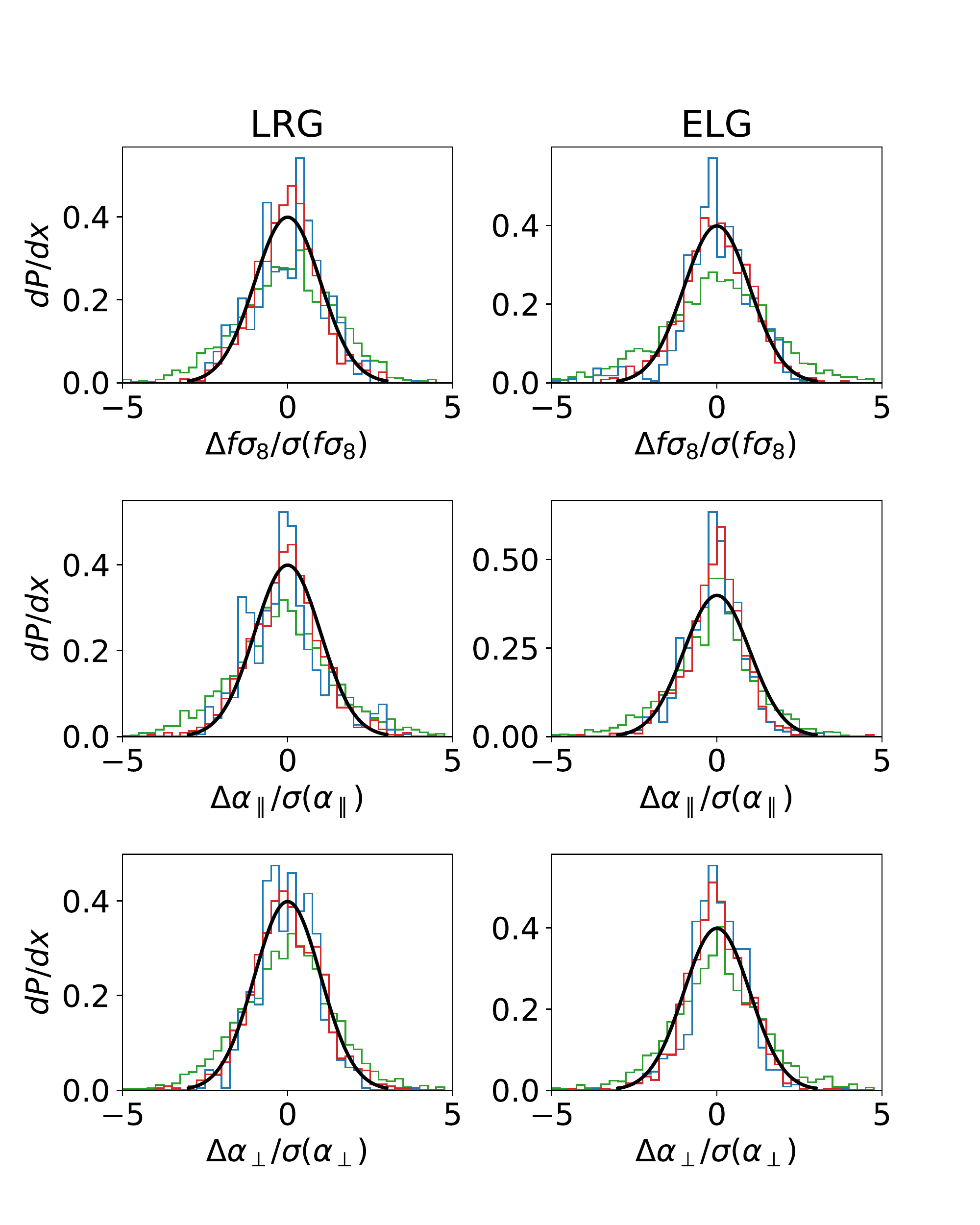} \vspace{-1 cm}
    \caption{Pull distributions for different covariance estimation techniques with results from fits on LRG and ELG mocks with line colors as in Fig. \ref{fig:fits_hists}}
    \label{fig:approx_hists}
\end{figure}

\begin{table}
\begin{center}
\begin{tabular}{||c c c c ||}
\hline
Survey & Mock & Mohammad-Percival & Fit \\
\hline
\hline
LRG & 1.04 & 1.49 & 1.07 \\
\hline
ELG & 1.08 & 1.80 & 1.07 \\
\hline
\end{tabular}
\end{center}
\caption{Standard deviation $\sigma$ of $(f\sigma_{8,i}-\overline{f\sigma_8})/\sigma_i(f\sigma_8)$, where $i$ is a separate fit for each of the methods. We can see, that for the mock covariance, it is close to 1 (as it is supposed to be when all of the fits share the same covariance.), for fitted covariance it is closing on it, and for jackknife usually takes values $>1.4$, which shows a much higher degree of deviation from what we assumed to be the truth.}
\label{tab:devs_EZ}
\end{table}

On Fig.~\ref{fig:approx_diags} the relative bias of the diagonals of jackknife-based vs mock-based covariances as defined by eq.~\ref{eqn:def_bias}  are shown for the LRG sample on the left and for the ELG sample on the right, in a similar way to Fig.~ \ref{fig:covs_mp}. First, The same trend is seen for the Mohammad-Percival jackknife as we found with the lognormal mocks: the bias of the jackknife method with the Mohammad-Percival correction tends to increase with number density, so from LRG to ELG, and the fitted jackknife is still able to mitigate it. However, we can also notice that the differences are less pronounced in the case of the EZmocks which is due to a bigger volume being probed by the same number density. In Appendix~\ref{sec:foot}, we test the impact of the size of the footprint on the diagonal elements of the covariance matrices by considering the North Galactic Cap, South Galactic Cap and full footprint separately.

As in the previous section, we also infer the values of the cosmological parameters $f\sigma_{8}$, $\alpha_{\parallel}$ and $\alpha_{\perp}$, using the same methodology as for the lognormal mocks. The results of the fits are shown in Fig.~\ref{fig:approx_fits} where the first row shows the $\chi^{2}/{\rm dof}$ distribution and the other rows show the marginalised 2D contours for best-fit values and uncertainties on the cosmological parameters. We confirm the findings with the lognormal mocks that the fitted jackknife method provides results which are in much better agreement with the mock-based method while the jackknife method with the Mohammad-Percival correction over-estimates clearly the uncertainties on all the cosmological parameters. The effect is also stronger as the number density of the galaxy sample increases.  
Moreover, as we have fewer mocks than for the tests with the lognormal mocks, we can notice that the fitted covariance based on 50 mocks actually produces smaller contours overall than the mock covariance which uses 1000 EZmocks. 

In Fig.~\ref{fig:approx_hists}, we show the pull distribution as defined by eq.~\ref{eq:pull} for the cosmological parameters and the standard deviations of the $f\sigma_{8}$ distribution, which is taken as an example, are presented on Table \ref{tab:devs_EZ}. The results are also similar to the ones obtained with the lognormal mocks: both the fitted jackknife and mock covariances produce a Gaussian shape with $\sigma = 1$,  while the standard deviation of the pull distribution obtained using the Mohammad-Percival correction for the jackknife method is larger ($\sigma$=1.5, 1.8 for LRG and ELG respectively). This quantitative test thus demonstrates that the fitted jackknife method performs better in estimating an unbiased and accurate covariance matrix for the two-point correlation function. 

Overall, throughout all of the tests for varying number densities, different types of mocks and number of fitted mocks, the fitted jackknife approach shows a considerable improvement over the correction for standard jackknife proposed by~\cite{Mohammad_Percival}. The fitted jackknife approach can achieve an unbiased estimate of the covariance matrix with similar precision to a mock-based covariance but with the major advantage of requiring a much smaller number of mocks.

%----------------------------------
% Section 4
%----------------------------------

\section{Conclusions}
\label{sec:conclusions}

Obtaining an accurate covariance matrix is a key ingredient for any cosmological analysis, and raises a significant challenge due to the limitations in computing power for mock-based methods or in the assumptions used in the analytical approaches. Additionally, as was shown in a series of reviews comparing different approximate methods, they still have problems reproducing exactly the results of more computationally intensive codes, especially in the non-linear regime (\citealt{cov_review1}, \citealt{cov_review2}, \citealt{cov_review3}). Some works also focused on decreasing the number of simulations needed to obtain a precise covariance matrix \citep{carpool}, for example combining the results from N-body and approximate simulations.

In this work we have attempted to tackle this challenge with the use of internal resampling methods. In Section~\ref{sec:intro-cov}, we review the basics of the jackknife formalism for two-point correlation function covariance estimation and perform a test on a toy model which confirms the improvement brought by a correction to the standard jackknife approach proposed by \cite{Mohammad_Percival}. Instead of using an analytically fixed correction to some terms that enter the jackknife covariance matrix, we propose to fit the correction to a mock-based covariance obtained from a small number of mocks. Moreover, we also noticed an unconstrained term in the different pairs that comprise the jackknife estimate of the covariance matrix, which we propose to account for by the same fitted jackknife procedure. In Section~\ref{sec:test-mocks}, we have tested this fitted jackknife covariance method and compared its performance with respect to the jackknife method with Mohammad-Percival correction and to a mock-based approach using lognormal mocks and approximate EZ mocks. We showed that the underestimation of the covariance obtained when using the Mohammad-Percival correction increases with galaxy number density while the fitted jackknife covariance remains unbiased. Performing the cosmological inference showed that the fitted jackknife covariance based on 50 mocks performs with the same accuracy as the covariance created from 1000-1500 mocks, both in terms of precision (unbiased constraints) and accuracy (similar uncertainties). There is also a significant decrease in computational power needed and we also stress that the method is simple to implement on top of the standard jackknife covariance computation. We provide a Python package that contains the implementation of the fitted jackknife method: https://github.com/theonefromnowhere/FitCov

Future work may include further tests of such a fitted jackknife covariance estimation technique when applied to scales smaller than $\sim 20 h^{-1}$Mpc. We plan to investigate the small scales in another work that aims at fitting the clustering of DESI Early Data with this method and mock-based covariances in order to estimate the galaxy-halo connection for different galaxy samples. A similar technique could also be developed in Fourier space, however, it would require a proper treatment of the window function effects when splitting the footprint into subsamples, together with a significant computational effort. We leave for future work the application of such techniques to other statistics, such as 3-point correlation function. Such a fitted jackknife covariance method can also be beneficial for multi-tracer analysis where it could accommodate all the degrees of freedom needed without requiring too many additional mocks. We plan to continue this work and apply the multi-tracer technique on the upcoming DESI Bright Galaxy Survey (\citealt{Zarrouk2021}, \citealt{Hahn2022}) whose high-density sampling make it a challenging test of the performance of the fitted jackknife covariance method.

\section*{Acknowledgements}

The authors acknowledge and are highly grateful for the fruitful discussions with Will Percival, Arnaud de Mattia and Michael Rashkovetskyi.
ST and PZ acknowledge the Fondation CFM pour la Recherche for their financial support. 
PN and SC acknowledge STFC funding ST/T000244/1 and ST/X001075/1.

This work used the DiRAC@Durham facility managed by the Institute for Computational Cosmology on behalf of the STFC DiRAC HPC Facility (www.dirac.ac.uk). The equipment was funded by BEIS capital funding via STFC capital grants ST/K00042X/1, ST/P002293/1, ST/R002371/1 and ST/S002502/1, Durham University and STFC operations grant ST/R000832/1. DiRAC is part of the National e-Infrastructure.
%The computations were performed on cosma and cosma6 clusters of the DiRAC facility, situated in Durham University.

This research is supported by the Director, Office of Science, Office of High Energy Physics of the U.S. Department of Energy under Contract No. DE–AC02–05CH11231, and by the National Energy Research Scientific Computing Center, a DOE Office of Science User Facility under the same contract; additional support for DESI is provided by the U.S. National Science Foundation, Division of Astronomical Sciences under Contract No. AST-0950945 to the NSF’s National Optical-Infrared Astronomy Research Laboratory; the Science and Technologies Facilities Council of the United Kingdom; the Gordon and Betty Moore Foundation; the Heising-Simons Foundation; the French Alternative Energies and Atomic Energy Commission (CEA); the National Council of Science and Technology of Mexico (CONACYT); the Ministry of Science and Innovation of Spain (MICINN), and by the DESI Member Institutions: \url{https://www.desi.lbl.gov/collaborating-institutions}.

The authors are honoured to be permitted to conduct scientific research on Iolkam Du’ag (Kitt Peak), a mountain with particular significance to the Tohono O’odham Nation.

%%%%%%%%%%%%%%%%%%%%%%%%%%%%%%%%%%%%%%%%%%%%%%%%%%
\section*{Data Availability}
The lognormal mocks can be easily reproduced using the public \texttt{MockFactory} code at \url{https://github.com/cosmodesi/mockfactory}.
The approximate EZmocks will be made public with upcoming DESI data releases. Data enabling the reproduction of the plots in this paper are published at \url{https://zenodo.org/record/7635683}.



\bibliographystyle{mnras}
\bibliography{example} % if your bibtex file is called example.bib

\begin{thebibliography}{}
\makeatletter
\relax
\def\mn@urlcharsother{\let\do\@makeother \do\$\do\&\do\#\do\^\do\_\do\%\do\~}
\def\mn@doi{\begingroup\mn@urlcharsother \@ifnextchar [ {\mn@doi@}
  {\mn@doi@[]}}
\def\mn@doi@[#1]#2{\def\@tempa{#1}\ifx\@tempa\@empty \href
  {http://dx.doi.org/#2} {doi:#2}\else \href {http://dx.doi.org/#2} {#1}\fi
  \endgroup}
\def\mn@eprint#1#2{\mn@eprint@#1:#2::\@nil}
\def\mn@eprint@arXiv#1{\href {http://arxiv.org/abs/#1} {{\tt arXiv:#1}}}
\def\mn@eprint@dblp#1{\href {http://dblp.uni-trier.de/rec/bibtex/#1.xml}
  {dblp:#1}}
\def\mn@eprint@#1:#2:#3:#4\@nil{\def\@tempa {#1}\def\@tempb {#2}\def\@tempc
  {#3}\ifx \@tempc \@empty \let \@tempc \@tempb \let \@tempb \@tempa \fi \ifx
  \@tempb \@empty \def\@tempb {arXiv}\fi \@ifundefined
  {mn@eprint@\@tempb}{\@tempb:\@tempc}{\expandafter \expandafter \csname
  mn@eprint@\@tempb\endcsname \expandafter{\@tempc}}}

\bibitem[\protect\citeauthoryear{Aghanim et~al.,}{Aghanim
  et~al.}{2020}]{planck2018}
Aghanim N.,  et~al., 2020, \mn@doi [Astronomy {\&} Astrophysics]
  {10.1051/0004-6361/201833910}, 641, A6

\bibitem[\protect\citeauthoryear{{Alcock} \& {Paczynski}}{{Alcock} \&
  {Paczynski}}{1979}]{AlcockPaczynski}
{Alcock} C.,  {Paczynski} B.,  1979, \mn@doi [\nat] {10.1038/281358a0}, \href
  {https://ui.adsabs.harvard.edu/abs/1979Natur.281..358A} {281, 358}

\bibitem[\protect\citeauthoryear{Balaguera-Antolínez, Kitaura,
  Pellejero-Ibáñez, Zhao  \& Abel}{Balaguera-Antolínez et~al.}{2018}]{bam}
Balaguera-Antolínez A.,  Kitaura F.-S.,  Pellejero-Ibáñez M.,  Zhao C.,
  Abel T.,  2018, \mn@doi [Monthly Notices of the Royal Astronomical Society:
  Letters] {10.1093/mnrasl/sly220}, 483, L58

\bibitem[\protect\citeauthoryear{Blas, Lesgourgues  \& Tram}{Blas
  et~al.}{2011}]{class}
Blas D.,  Lesgourgues J.,   Tram T.,  2011, \mn@doi [Journal of Cosmology and
  Astroparticle Physics] {10.1088/1475-7516/2011/07/034}, 2011, 034

\bibitem[\protect\citeauthoryear{Blot et~al.,}{Blot et~al.}{2019}]{cov_review2}
Blot L.,  et~al., 2019, \mn@doi [Monthly Notices of the Royal Astronomical
  Society] {10.1093/mnras/stz507}, 485, 2806

\bibitem[\protect\citeauthoryear{Chartier, Wandelt, Akrami  \&
  Villaescusa-Navarro}{Chartier et~al.}{2021}]{carpool}
Chartier N.,  Wandelt B.,  Akrami Y.,   Villaescusa-Navarro F.,  2021, \mn@doi
  [Monthly Notices of the Royal Astronomical Society] {10.1093/mnras/stab430},
  503, 1897

\bibitem[\protect\citeauthoryear{Chen, Vlah  \& White}{Chen
  et~al.}{2020}]{velocileptors1}
Chen S.-F.,  Vlah Z.,   White M.,  2020, \mn@doi [Journal of Cosmology and
  Astroparticle Physics] {10.1088/1475-7516/2020/07/062}, 2020, 062

\bibitem[\protect\citeauthoryear{Chen, Vlah, Castorina  \& White}{Chen
  et~al.}{2021}]{velocileptors2}
Chen S.-F.,  Vlah Z.,  Castorina E.,   White M.,  2021, \mn@doi [Journal of
  Cosmology and Astroparticle Physics] {10.1088/1475-7516/2021/03/100}, 2021,
  100

\bibitem[\protect\citeauthoryear{Chuang, Kitaura, Prada, Zhao  \& Yepes}{Chuang
  et~al.}{2014}]{ez_mocks}
Chuang C.-H.,  Kitaura F.-S.,  Prada F.,  Zhao C.,   Yepes G.,  2014, \mn@doi
  [Monthly Notices of the Royal Astronomical Society] {10.1093/mnras/stu2301},
  446, 2621

\bibitem[\protect\citeauthoryear{Colavincenzo et~al.,}{Colavincenzo
  et~al.}{2018}]{cov_review3}
Colavincenzo M.,  et~al., 2018, \mn@doi [Monthly Notices of the Royal
  Astronomical Society] {10.1093/mnras/sty2964}, 482, 4883

\bibitem[\protect\citeauthoryear{{DESI Collaboration} et~al.,}{{DESI
  Collaboration} et~al.}{2016}]{desi}
{DESI Collaboration} et~al., 2016, arXiv e-prints, \href
  {https://ui.adsabs.harvard.edu/abs/2016arXiv161100036D} {p. arXiv:1611.00036}

\bibitem[\protect\citeauthoryear{{DESI Collaboration} et~al.,}{{DESI
  Collaboration} et~al.}{2022}]{DESI2022}
{DESI Collaboration} et~al., 2022, \mn@doi [\aj] {10.3847/1538-3881/ac882b},
  \href {https://ui.adsabs.harvard.edu/abs/2022AJ....164..207A} {164, 207}

\bibitem[\protect\citeauthoryear{{Dawson} et~al.,}{{Dawson}
  et~al.}{2013}]{Dawson2013}
{Dawson} K.~S.,  et~al., 2013, \mn@doi [\aj] {10.1088/0004-6256/145/1/10},
  \href {https://ui.adsabs.harvard.edu/abs/2013AJ....145...10D} {145, 10}

\bibitem[\protect\citeauthoryear{Dembinski \& et al.}{Dembinski \&
  et~al.}{2020}]{iminuit}
Dembinski H.,  et al. P.~O.,  2020, \mn@doi [] {10.5281/zenodo.3949207}

\bibitem[\protect\citeauthoryear{Favole, Granett, Silva Lafaurie  \&
  Sapone}{Favole et~al.}{2021}]{Favelo2020}
Favole G.,  Granett B.~R.,  Silva Lafaurie J.,   Sapone D.,  2021, \mn@doi
  [Monthly Notices of the Royal Astronomical Society] {10.1093/mnras/stab1720},
  505, 5833

\bibitem[\protect\citeauthoryear{{Feldman}, {Kaiser}  \& {Peacock}}{{Feldman}
  et~al.}{1994}]{FKP}
{Feldman} H.~A.,  {Kaiser} N.,   {Peacock} J.~A.,  1994, \mn@doi [\apj]
  {10.1086/174036}, \href
  {https://ui.adsabs.harvard.edu/abs/1994ApJ...426...23F} {426, 23}

\bibitem[\protect\citeauthoryear{Feng, Chu, Seljak  \& McDonald}{Feng
  et~al.}{2016}]{fast_pm}
Feng Y.,  Chu M.-Y.,  Seljak U.,   McDonald P.,  2016, \mn@doi [Monthly Notices
  of the Royal Astronomical Society] {10.1093/mnras/stw2123}, 463, 2273

\bibitem[\protect\citeauthoryear{{Friedrich}, {Seitz}, {Eifler}  \&
  {Gruen}}{{Friedrich} et~al.}{2016}]{Friedrich2016}
{Friedrich} O.,  {Seitz} S.,  {Eifler} T.~F.,   {Gruen} D.,  2016, \mn@doi
  [\mnras] {10.1093/mnras/stv2833}, \href
  {https://ui.adsabs.harvard.edu/abs/2016MNRAS.456.2662F} {456, 2662}

\bibitem[\protect\citeauthoryear{Hahn et~al.,}{Hahn et~al.}{2022}]{Hahn2022}
Hahn C.,  et~al., 2022, DESI Bright Galaxy Survey: Final Target Selection,
  Design, and Validation, \mn@doi{10.48550/ARXIV.2208.08512}, \url
  {https://arxiv.org/abs/2208.08512}

\bibitem[\protect\citeauthoryear{{Hartlap, J.}, {Simon, P.}  \& {Schneider,
  P.}}{{Hartlap, J.} et~al.}{2007}]{Hartlap}
{Hartlap, J.} {Simon, P.}  {Schneider, P.} 2007, \mn@doi [A\&A]
  {10.1051/0004-6361:20066170}, 464, 399

\bibitem[\protect\citeauthoryear{{Ivezi{\'c}} et~al.,}{{Ivezi{\'c}}
  et~al.}{2019}]{lsst}
{Ivezi{\'c}} {\v Z}.,  et~al., 2019, \mn@doi [\apj] {10.3847/1538-4357/ab042c},
  \href {http://adsabs.harvard.edu/abs/2019ApJ...873..111I} {873, 111}

\bibitem[\protect\citeauthoryear{Kitaura et~al.,}{Kitaura
  et~al.}{2016}]{patchy}
Kitaura F.-S.,  et~al., 2016, \mn@doi [Monthly Notices of the Royal
  Astronomical Society] {10.1093/mnras/stv2826}, 456, 4156

\bibitem[\protect\citeauthoryear{Klypin \& Prada}{Klypin \& Prada}{2018}]{glam}
Klypin A.,  Prada F.,  2018, \mn@doi [Monthly Notices of the Royal Astronomical
  Society] {10.1093/mnras/sty1340}, 478, 4602

\bibitem[\protect\citeauthoryear{{Lacasa, Fabien} \& {Kunz, Martin}}{{Lacasa,
  Fabien} \& {Kunz, Martin}}{2017}]{Lacasa2017}
{Lacasa, Fabien} {Kunz, Martin} 2017, \mn@doi [A\&A]
  {10.1051/0004-6361/201730784}, 604, A104

\bibitem[\protect\citeauthoryear{{Landy} \& {Szalay}}{{Landy} \&
  {Szalay}}{1993}]{LandySzalay}
{Landy} S.~D.,  {Szalay} A.~S.,  1993, \mn@doi [\apj] {10.1086/172900}, \href
  {https://ui.adsabs.harvard.edu/abs/1993ApJ...412...64L} {412, 64}

\bibitem[\protect\citeauthoryear{Laureijs et~al.,}{Laureijs
  et~al.}{2011}]{Laureijs2011}
Laureijs R.,  et~al., 2011, Euclid Definition Study Report,
  \mn@doi{10.48550/ARXIV.1110.3193}, \url {https://arxiv.org/abs/1110.3193}

\bibitem[\protect\citeauthoryear{Lippich et~al.,}{Lippich
  et~al.}{2018}]{cov_review1}
Lippich M.,  et~al., 2018, \mn@doi [Monthly Notices of the Royal Astronomical
  Society] {10.1093/mnras/sty2757}, 482, 1786

\bibitem[\protect\citeauthoryear{Maksimova, Garrison, Eisenstein, Hadzhiyska,
  Bose  \& Satterthwaite}{Maksimova et~al.}{2021}]{abacus}
Maksimova N.~A.,  Garrison L.~H.,  Eisenstein D.~J.,  Hadzhiyska B.,  Bose S.,
   Satterthwaite T.~P.,  2021, \mn@doi [Monthly Notices of the Royal
  Astronomical Society] {10.1093/mnras/stab2484}, 508, 4017

\bibitem[\protect\citeauthoryear{Mohammad \& Percival}{Mohammad \&
  Percival}{2022}]{Mohammad_Percival}
Mohammad F.~G.,  Percival W.~J.,  2022, \mn@doi [Monthly Notices of the Royal
  Astronomical Society] {10.1093/mnras/stac1458}, 514, 1289

\bibitem[\protect\citeauthoryear{Norberg, Baugh, Gaztañaga  \& Croton}{Norberg
  et~al.}{2009}]{Nordberg}
Norberg P.,  Baugh C.~M.,  Gaztañaga E.,   Croton D.~J.,  2009, \mn@doi
  [Monthly Notices of the Royal Astronomical Society]
  {10.1111/j.1365-2966.2009.14389.x}, 396, 19

\bibitem[\protect\citeauthoryear{O'Connell, Eisenstein, Vargas, Ho  \&
  Padmanabhan}{O'Connell et~al.}{2016}]{RascalC2}
O'Connell R.,  Eisenstein D.,  Vargas M.,  Ho S.,   Padmanabhan N.,  2016,
  \mn@doi [Monthly Notices of the Royal Astronomical Society]
  {10.1093/mnras/stw1821}, 462, 2681

\bibitem[\protect\citeauthoryear{{Percival} et~al.,}{{Percival}
  et~al.}{2014}]{Percival2014}
{Percival} W.~J.,  et~al., 2014, \mn@doi [\mnras] {10.1093/mnras/stu112}, \href
  {https://ui.adsabs.harvard.edu/abs/2014MNRAS.439.2531P} {439, 2531}

\bibitem[\protect\citeauthoryear{Philcox, Eisenstein, O’Connell  \&
  Wiegand}{Philcox et~al.}{2019}]{RascalC1}
Philcox O. H.~E.,  Eisenstein D.~J.,  O’Connell R.,   Wiegand A.,  2019,
  \mn@doi [Monthly Notices of the Royal Astronomical Society]
  {10.1093/mnras/stz3218}, 491, 3290

\bibitem[\protect\citeauthoryear{Taruya, Nishimichi  \& Saito}{Taruya
  et~al.}{2010}]{Taruya}
Taruya A.,  Nishimichi T.,   Saito S.,  2010, \mn@doi [Phys. Rev. D]
  {10.1103/PhysRevD.82.063522}, 82, 063522

\bibitem[\protect\citeauthoryear{Tassev, Zaldarriaga  \& Eisenstein}{Tassev
  et~al.}{2013}]{cola}
Tassev S.,  Zaldarriaga M.,   Eisenstein D.~J.,  2013, \mn@doi [Journal of
  Cosmology and Astroparticle Physics] {10.1088/1475-7516/2013/06/036}, 2013,
  036

\bibitem[\protect\citeauthoryear{{Wadekar}, {Ivanov}  \&
  {Scoccimarro}}{{Wadekar} et~al.}{2020}]{Wadekar1}
{Wadekar} D.,  {Ivanov} M.~M.,   {Scoccimarro} R.,  2020, \mn@doi [\prd]
  {10.1103/PhysRevD.102.123521}, \href
  {https://ui.adsabs.harvard.edu/abs/2020PhRvD.102l3521W} {102, 123521}

\bibitem[\protect\citeauthoryear{Wishart}{Wishart}{1928}]{Wishart1928}
Wishart J.,  1928, \mn@doi [Biometrika] {10.1093/biomet/20A.1-2.32}, 20A, 32

\bibitem[\protect\citeauthoryear{Zarrouk et~al.,}{Zarrouk
  et~al.}{2021}]{Zarrouk2021}
Zarrouk P.,  et~al., 2021, \mn@doi [Monthly Notices of the Royal Astronomical
  Society] {10.1093/mnras/stab2814}, 509, 1478

\bibitem[\protect\citeauthoryear{{Zel'dovich}}{{Zel'dovich}}{1970}]{Zeldovich}
{Zel'dovich} Y.~B.,  1970, \aap, \href
  {https://ui.adsabs.harvard.edu/abs/1970A&A.....5...84Z} {5, 84}

\bibitem[\protect\citeauthoryear{Zhao et~al.,}{Zhao
  et~al.}{2021}]{ez_mocks_eboss}
Zhao C.,  et~al., 2021, \mn@doi [Monthly Notices of the Royal Astronomical
  Society] {10.1093/mnras/stab510}, 503, 1149

\makeatother
\end{thebibliography}

%\begin{thebibliography}{99}
%\end{thebibliography}

%%%%%%%%%%%%%%%%%%%%%%%%%%%%%%%%%%%%%%%%%%%%%%%%%%

%%%%%%%%%%%%%%%%% APPENDICES %%%%%%%%%%%%%%%%%%%%%

\appendix
\section{Impact of the size of the footprint}
\label{sec:foot}

The lognormal and approximate EZmocks used in this work for testing the different covariance matrix estimates do not use the same size  survey footprint. Given that 
 for the same number density we see a bigger discrepancy between the jackknife method with Mohammad-Percival correction and the mock-based covariance matrix in the case of the lognormal mocks than with the approximate mocks, we also explore the effect of varying the size of the footprint. We consider the LRG EZmocks and compute the covariance matrix for the three methods (jackknife with Mohammad-Percival, fitted jackknife and mock-based) for the Southern Galactic Cap  separately and compare the results with the ones for the full Y5 footprint. We keep the same number of jackknife regions in all cases. 

\begin{figure}
    \centering
    \includegraphics[width=1.\linewidth]{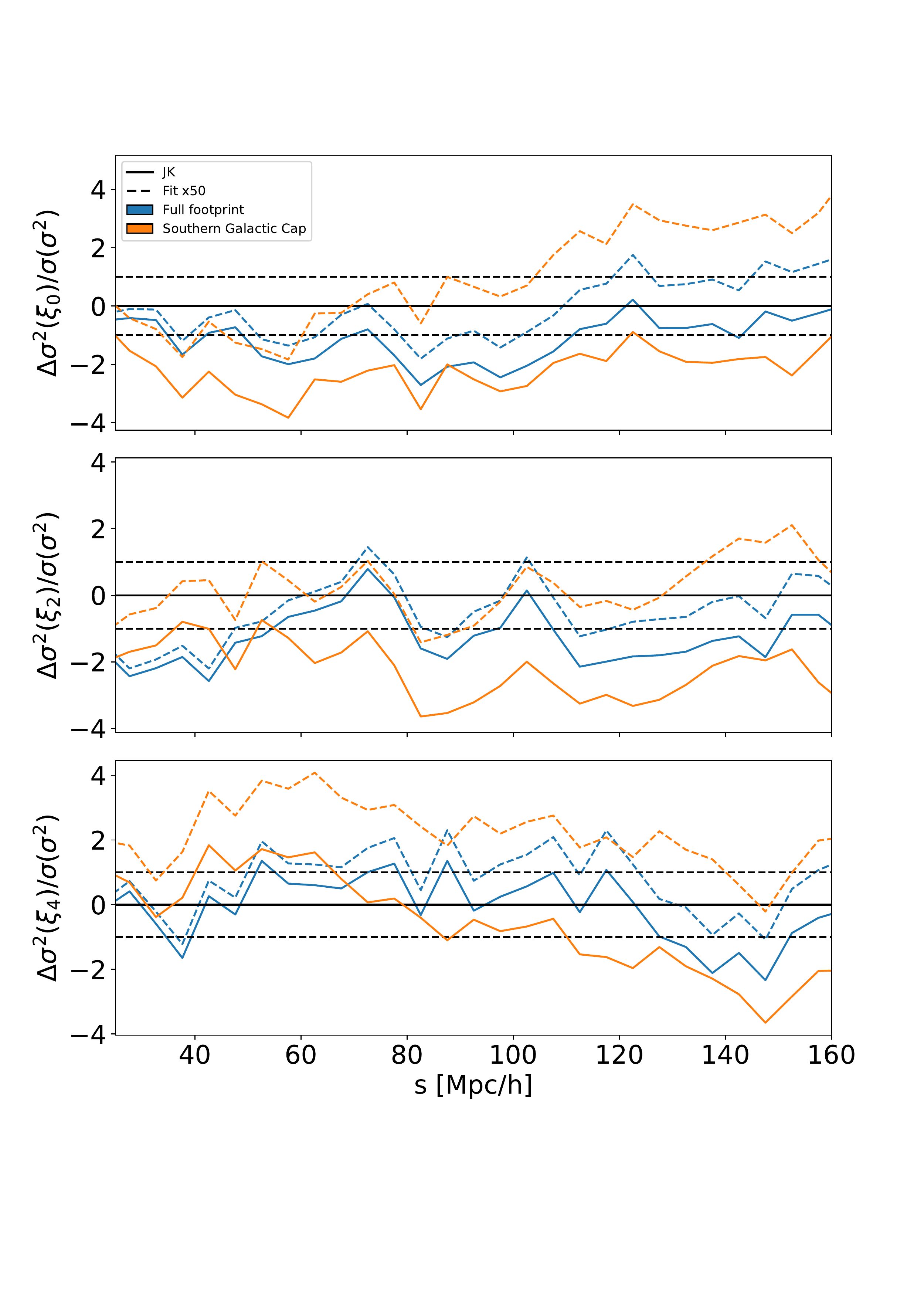}\vspace{-2.2 cm}
    \caption{Relative bias as defined in equation (\ref{eqn:def_bias}) for the fitted jackknife method (dashed) and for the Mohammad-Percival approach (solid) as a function of pair separation for LRG EZ mocks. The orange curves are the results for the SGC while the blue curves are for the larger full footprint.}
    \label{fig:footprint_test}
\end{figure}

The results are displayed in Fig. \ref{fig:footprint_test} where we plot the relative bias as defined by equation (\ref{eqn:def_bias}) between a jackknife method and the mock-based covariance as a function of pair separation for the monopole (top), quadrupole (middle) and hexadecapole (bottom). The results for the SGC are shown in orange and the ones for the full footprint in blue. Indeed, we see that the bias associated with the Mohammad-Percival approach is higher when a smaller footprint is used. We also confirm the same effect for the ELG dataset. Therefore, it makes the fitted jackknife covariance method even more useful when the footprint considered is relatively small. We believe that this effect might be related to super-sampling covariance, and if it is indeed the case, it would mean that our approach can successfully take it into account.

%%%%%%%%%%%%%%%%%%%%%%%%%%%%%%%%%%%%%%%%%%%%%%%%%%

% Don't change these lines
\bsp	% typesetting comment
\label{lastpage}
\end{document}